\documentclass[12pt,preprint]{emulateapj}
\usepackage{apjfonts}
\usepackage{amsmath}
\usepackage{epsf}
\usepackage{color}
\usepackage{color}

\shorttitle{Increasing Stellar Baryon Fractions at $z >$ 4}
\shortauthors{Finkelstein et al.}

\newcommand{\sol}{$_{\odot}$}

\begin{document}
\slugcomment{Accepted to the Astrophysical Journal}
\title{An Increasing Stellar Baryon Fraction in Bright Galaxies at
  High Redshift}

\author{Steven L. Finkelstein\altaffilmark{1,a}, Mimi
  Song\altaffilmark{1}, Peter Behroozi\altaffilmark{2},  Rachel S.\
  Somerville\altaffilmark{3}, Casey Papovich\altaffilmark{4}, Milo\v s
  Milosavljevi\'c\altaffilmark{1}, Avishai Dekel\altaffilmark{5}, Desika
  Narayanan\altaffilmark{6}, Matthew L. N. Ashby\altaffilmark{7},
  Asantha Cooray\altaffilmark{8}, 
Giovanni G.\ Fazio\altaffilmark{7}, Henry C.\ Ferguson\altaffilmark{2},
Anton M.\ Koekemoer\altaffilmark{2}, Brett Salmon\altaffilmark{4}, \& S. P. Willner\altaffilmark{7}}
\altaffiltext{1}{Department of Astronomy, The University of Texas at Austin, Austin, TX 78712}
\altaffiltext{2}{Space Telescope Science Institute, Baltimore, MD 21218}
\altaffiltext{3}{Department of Physics \& Astronomy, Rutgers University, 136 Frelinghuysen Road, Piscataway, NJ 08854}
\altaffiltext{4}{George P. and Cynthia Woods Mitchell Institute for
  Fundamental Physics and Astronomy, Department of Physics and
  Astronomy, Texas A\&M University, College Station, TX 77843}
\altaffiltext{5}{Racah Institute of Physics, The Hebrew University, Jerusalem 91904, Israel}
\altaffiltext{6}{Department of Physics and Astronomy, Haverford College, Haverford, PA 19041}
\altaffiltext{7}{Harvard-Smithsonian Center for Astrophysics, 60 Garden Street, Cambridge, Massachusetts 02138}
\altaffiltext{8}{Center for Cosmology, Department of Physics and Astronomy, University of California, Irvine, CA 92697}
\altaffiltext{a}{stevenf@astro.as.utexas.edu}

\begin{abstract}
Recent observations have shown that the characteristic
luminosity of the rest-frame ultraviolet (UV) luminosity function does
not significantly evolve at $4 < z < 7$ and is approximately
$M^{\ast}_\mathrm{UV}\sim-21$.  We investigate this apparent
non-evolution by examining a sample of 178 bright, $M_\mathrm{UV}
< -$21 galaxies at $z =$ 4 to 7, analyzing their 
stellar populations and host halo masses.
Including deep {\it Spitzer}/IRAC imaging
to constrain the rest-frame optical light, we find that
$M^{\ast}_\mathrm{UV}$ galaxies at $z =$ 4--7 have similar stellar masses of $\log(M/M_\odot) =$
9.6--9.9 and are thus relatively massive for these high redshifts. 
However, bright galaxies at $z =$ 4--7 are less massive and have younger inferred ages than
similarly bright galaxies at $z =$ 2--3, even though the two populations have
similar star formation rates and levels of dust attenuation for a
fixed dust-attenuation curve.  Matching
the abundances of these bright $z =$ 4--7 galaxies to halo mass
functions from the Bolshoi $\Lambda$CDM simulation implies that the
typical halo masses in $\sim M^{\ast}_{\rm UV}$
galaxies decrease from $\log(M_{\rm h}/M_\odot) =11.9$ at $z =$ 4 to
$\log(M_{\rm h}/M_\odot) = 11.4$ at $z =7$.  Thus, although we are
studying galaxies at a similar mass across multiple redshifts, these
galaxies live in lower mass halos at higher redshift.
The stellar baryon fraction in units of the cosmic mean $\Omega_{\rm
  b}/\Omega_{\rm m}$ rises from 5.1\% at $z =4$ to 11.7\% at $z =7$; this
evolution is significant at the
$\sim3\sigma$ level.  This rise does not agree with simple expectations
of how galaxies grow, and implies that some effect, perhaps a diminishing efficiency
of feedback, is allowing a higher fraction of available baryons to be
converted into stars at high redshifts.
\end{abstract}


\begin{figure*}[!t]
\epsscale{1.15}
\plotone{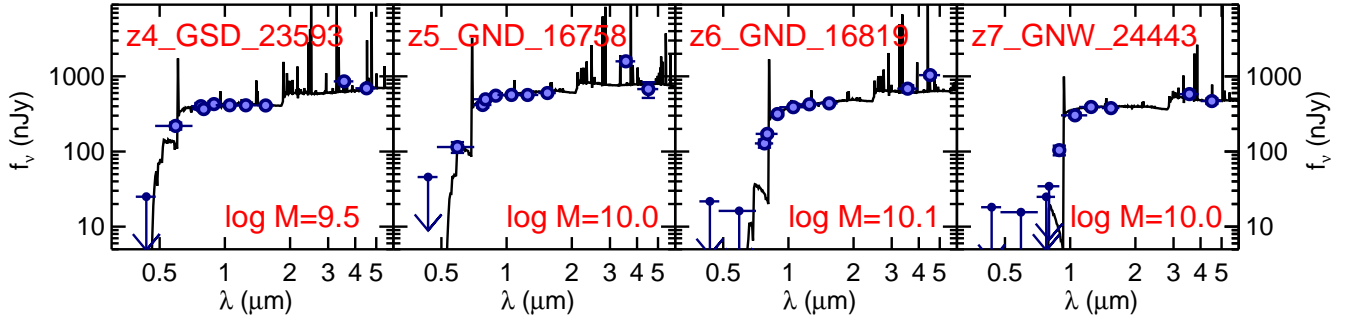}
\caption{The spectral energy distribution of a representative galaxy in each of our
  redshift bins, shown as the blue circles.  For filters where the
  signal-to-noise was $<$2, we show the 1$\sigma$ upper
  limits.  The black curve shows the best-fit stellar population
  model, and we list the ID and best-fit stellar mass for each galaxy.}
\label{fig:meansed}
\end{figure*}

\section{Introduction}

Tracing the buildup of stellar mass from the epoch of the first
galaxies through the present can be used to constrain
models of galaxy formation.  Understanding what physics governs this buildup is one of
the key outstanding questions in galaxy evolution.  There is a consensus that the cosmic
star formation rate (SFR) density rises from the dawn of galaxies, peaks at redshifts $z
\sim$ 2--3, and then declines steeply at $z <$ 2 \citep[e.g.,][]{madau14}. The physical origin of this
evolution in galaxy stellar mass growth is poorly understood, especially at high redshifts.  
While studies of galaxy evolution
routinely quantify the stellar content of distant galaxies, it remains
challenging to relate the stellar masses to the supply of gas fueling
star formation.  Theoretical works attempt to address these fundamental
questions to a varying degree of success, but observational data have
remained incomplete, particularly at the massive end, and they are dominated by systematic uncertainties
unavoidable in the stitching together of datasets from
different campaigns (e.g., Behroozi et al.\ 2013a).  

The advances facilitated by the Wide Field Camera 3 (WFC3) on the {\it Hubble Space
  Telescope} ({\it HST}) over the past half-decade have led to the discovery
of $>$1000 redshift $z >$ 6 galaxies
\citep[e.g.,][]{finkelstein10,finkelstein12b,finkelstein12a,finkelstein13,finkelstein15,bouwens10a,bouwens12,bouwens14,bouwens15,oesch10,oesch12,oesch13,oesch14,mclure10,mclure13,wilkins11b,schenker13}.
Among the detailed analyses facilitated by these large samples is the
measurement of the rest-frame ultraviolet luminosity function,
which quantifies the
relative abundances of galaxies over a wide dynamic range in
luminosity.  As the UV light probes recent star formation
activity, the integral of the rest-frame UV luminosity function provides an
estimate of the cosmic SFR density
\citep[e.g.,][]{madau96,bouwens12,madau14,finkelstein15}.
The luminosity function is typically parameterized with the
\citet{schechter76} functional form that is a power law  at low
luminosities and declines exponentially at high luminosities.
Its parameters are the characteristic luminosity
$M^{\ast}_{\rm UV}$, the faint-end slope $\alpha$, and the normalization
$\phi^{\ast}$.  Previous studies typically found that these parameters
evolved with redshift: the characteristic luminosity decreased with
increasing redshift as the faint-end slope steepened
\citep[e.g.,][]{bouwens07,bouwens11d,bouwens11,mclure13}.  This
``luminosity evolution'' of the luminosity function was widely
accepted as it fit the
general trend observable in the evolution of the cosmic SFR density.

More recent work, however, has shown that the picture described above is incomplete.
The first evidence came from \citet{ono12}, \citet{finkelstein13}, and
\citet{bowler14}, where a larger than expected number of
bright galaxies turned up as $z =$ 7 surveys moved to wider
fields.  \citet{finkelstein15} and \citet{bouwens15} have confirmed
this excess, and recent studies have concluded it is not attributable to gravitational
lensing \citep{mason15,barone-nugent15}.  Both \citet{finkelstein15} and \citet{bouwens15} computed the evolution
of the rest-frame UV luminosity function at $z =$ 4--8, finding that
contrary to the preceding results derived from smaller datasets, the characteristic
luminosity $M^{\ast}_{\rm UV}$ was remarkably redshift-independent between $z =$ 4, 5, 6, and 7.   The constancy broke down only at $z =$ 8 where the data were least
constraining.  Most of the evolution took 
place in the characteristic number density: it declined towards higher redshifts.
Therefore, while galaxies in general became less common at higher
redshifts---consistent with the decline in the cosmic SFR density---bright galaxies
remained relatively common in the distant universe.

Here, we seek to constrain the physical properties in distant
UV-bright galaxies and attempt to understand how they maintained high
levels of star formation.  In \S 2, we describe the bright galaxy sample that we have taken
from \citet{finkelstein15} and discuss the additional constraints that can be
placed on the stellar populations with the inclusion of
{\it Spitzer}/IRAC photometry.  In \S 3, we use cosmic abundance matching to estimate the 
halo masses and the stellar-to-halo mass ratios for these galaxies.  In \S 4, we discuss the
evolution of the stellar baryon fraction with redshift, and in \S 5, we
present our conclusions. We
assume the WMAP7 $\Lambda$CDM cosmological model \citep{komatsu11} throughout, with $H_0 = 70.2$
km s$^{-1}$ Mpc$^{-1}$, $\Omega_{\rm m} =$ 0.275, and $\Omega_{\Lambda} =$ 0.725.
All magnitudes given are in the AB system
\citep{oke83}.

\section{Stellar Populations in UV-Bright Galaxies}

Here we wish to constrain the physical processes that regulate the
abundance of bright galaxies in the distant universe.  These luminous
systems are observed only a short time after the Big Bang and trace prominent 
density peaks at their epoch.  We further investigate
sources with $M_{1500} < -$21.  This magnitude is approximately the value of
$M^{\ast}_{\rm UV}$ at these redshifts
\citep{bouwens15,finkelstein15}, though the exact value
of $M^{\ast}_{\rm UV}$ does get progressively more uncertain with
increasing redshift ($\pm$ 0.09 at $z =$ 4 to $\pm$ 0.4 at $z =$ 7; \citealt{finkelstein15}).
Using ground-based data, \citet{bowler15} have found evidence
that $M^{\ast}_{\rm UV}$ may be fainter than $-$21 at $z =$ 7, though only at the
2$\sigma$ level, and thus not significantly discrepant with the
measurements of \citet{bouwens15} and \citet{finkelstein15}.  Because
this study is concerned with the physics driving the apparently
high star-formation rates in distant bright galaxies, the exact
luminosity we choose is not critical.  In order to explore relatively bright galaxies, we choose the
approximate value of $M^{\ast}_{\rm UV}$ at these redshifts as our threshold.

We use the sample of $\sim7500$
galaxies at $4\lesssim z \lesssim8$  from \citet{finkelstein15} and refer the reader to that paper
for details on the photometry, photometric redshift
sample selection, and derivation of UV absolute magnitude at 1500 \AA\
($M_\mathrm{UV}$).  From their full catalog, here we analyze the
150, 75, 28, 18, and 3 galaxies with $M_\mathrm{UV} < -$21 at $z =$ 4,
5, 6, 7, and 8, respectively, which come from the CANDELS GOODS-South
and North fields (with a total area of $\sim$ 280 arcmin$^2$). The redshift bins are bounded by $\Delta
z=\pm 0.5$ from the central redshift. 

\subsection{Inclusion of {\it Spitzer}/IRAC Photometry}

 To learn more about these intriguing systems,
we turn to stellar population modeling.  We use the {\it HST}
photometry from \citet{finkelstein15} that includes ACS and WFC3
PSF-matched total fluxes in the wavelength range 0.4 -- 1.8 $\mu$m
\citep[see][for details on the imaging]{koekemoer11}.  
This catalog also includes photometry from the {\it Spitzer Space Telescope} Infrared
Array Camera \citep[IRAC;][]{fazio04} imaging of our fields.  The
IRAC imaging probes the rest-frame optical at $z >$ 4 and
thus provides significant constraining power on the stellar masses of
our galaxies.  It also gives a more accurate handle on the ages and dust
attenuations by reducing the degeneracy between the two parameters.
The details of the long-wavelength photometry are presented in
\citet{finkelstein15}; here, we review them only briefly.  The mosaics
were obtained by coadding all the available data in these fields:
the GOODS (Dickinson et al., in prep), {\it Spitzer} Extended Deep
Survey \citep[SEDS;][]{ashby13}, and {\sl Spitzer}-CANDELS
\citep[S-CANDELS;][]{ashby15} wide-field programs, as well as the deep
pointings from {\it Spitzer} program 70145 \citep[the
IRAC Ultra-Deep Field of][]{labbe13} over the Hubble Ultra Deep Field
and its parallels, and program 70204 (PI Fazio) which observed a
region in the GOODS-S field to 100 hr depth.  The final mosaics have a
depth of $\gtrsim50$ hr over both CANDELS GOODS
fields and $>$100 hr over the HUDF main field \citep{ashby15}.

The {\tt TPHOT} software \citep{merlin15}, an updated version
of {\tt TFIT} (Laidler et al. 2007), was used to measure photometry in
the {\it Spitzer}/IRAC imaging.  This software models the
low-resolution IRAC images by convolving the {\it HST}/WFC3 $H$-band
image with an empirically derived IRAC PSF, simultaneously fitting all
IRAC sources.  This provides robust photometry even for moderately
blended sources.  The full description of our {\tt TPHOT} IRAC
photometry catalog is presented in \citet{song15}.

All high-redshift galaxies were visually inspected
in the TPHOT residual maps.   If an object was on or near a strong
residual, reliable IRAC photometry was not possible, affecting
20-30\% of the galaxies in our bright galaxy sample.
To obtain the most robust stellar mass measurements, in our subsequent
analysis we do not include these affected galaxies.
Over 90\% of the remaining galaxies in our sample had a 3.6 $\mu$m or
4.5 $\mu$m detection of at least 3$\sigma$ significance, with a magnitude range
at $z \geq$ 6 of $23.5 \leq m_{3.6} \leq 25.5$.  This is expected, as
galaxies with $M_{1500} < -21$ should be massive enough at all
redshifts to yield an IRAC detection absent crowding.  In fact, when comparing median
stellar masses derived excluding and including galaxies without IRAC constraints, the
median stellar mass of galaxies with IRAC constraints is at most $\sim0.1$ dex
\emph{lower} than that of the whole sample.  This is because
galaxies with true mid-infrared fluxes well below the IRAC detection
limit can have poorly constrained SEDs.  In contrast, our results show that the typical UV
  bright $z >$ 6 galaxy has a lower mass-to-light ratio than other possible solutions,
thus the IRAC detection prior does not drive us to higher M/L models.
 
The IRAC photometry used here is the same as by
\citet{finkelstein15} who used the IRAC and
additional {\it HST}/ACS F814W photometry to re-measure the
photometric redshifts, removing 14, 14, and 1 galaxies from their $z =$
4, 5, and 6 samples, respectively.  The galaxy sample we
consider here comes from the cleaned sample of \citet{finkelstein15},
thus these presumed lower redshift interlopers have already been
removed.  In this work, we have performed an additional iteration of
visual inspection of the IRAC imaging, which results in a few more
galaxies having preferred lower redshift solutions.  From our sample we remove
seven such additional galaxies which
now have preferred lower redshift solutions (two from our $z=$4
sample, and five from our $z=$5 sample).  We also remove five
additional sources from our $z =$ 4 sample which have 2.7 $<
z_\mathrm{phot} <$ 3.5, as we do not wish to bias our $z =$ 4 sample
stellar mass measurements.
After excluding these objects, galaxies in the remaining sample all
have photometric redshifts within $z_\mathrm{sample} - 0.5 <
z_\mathrm{phot} < z_\mathrm{sample} + 0.7$, thus we consider the
effects of potential low-redshift interlopers to be minimal.  The
inclusion of the very small number of sources at $z_\mathrm{phot} =
z_\mathrm{sample} +$ 0.5 -- 0.7 (eight at $z =$ 4, six
at $z =$ 5 and 1 at $z =$ 7; see tables in Appendix) make no difference
to the median stellar mass discussed below.  See
\citet{finkelstein15} for a quantitative description of the potential
contamination.

\subsection{Spectral Energy Distribution Fitting}

The technique we used to fit stellar population models to 
photometry was similar to the one we employed before
\citep{finkelstein10,finkelstein12a,finkelstein12b,finkelstein13,finkelstein15}.  We
used the updated (2007) stellar population synthesis
models of \citet{bruzual03} to generate a grid of model spectra\footnote[1]{The 2007
  update to the stellar population models of \citet{bruzual03} may overestimate the
  contribution of thermally pulsating asymptotic giant branch (TP-AGB)
stars.  However, these stars typically begin to dominate the emission at
population ages $\gtrsim$1 Gyr and rest frame wavelengths
$\gtrsim 1 \,\mu$m.  Our longest wavelength filter (4.5 $\mu$m) at our
lowest redshift ($z =$ 4) probes only 0.9 $\mu$m, and all other
filter/redshift combinations probe bluer rest-frame wavelengths.
The TP-AGB contribution may impact the SED in
post-starburst galaxies at wavelengths as low as 0.5 $\mu$m \citep{kriek10}.
However, our galaxies are highly star-forming, with inferred population ages are $\ll$1 Gyr.  Thus, 
our choice to use the updated models likely has no effect on our results.}.  We varied
the stellar mass (defined as the total gas mass converted into stars), the
stellar population metallicity, the time since the onset of star
formation (henceforth, the age), and the star formation history (SFH). We 
assumed the Salpeter\footnote[2]{To convert our results to those
  obtained from a Chabrier IMF, one should divide the stellar masses
  and SFRs by a factor of $\sim1.7$.} initial mass function (IMF).
Allowed metallicities spanned (0.02 -- 1)\,$Z_\odot$ and ages spanned 1 Myr to
the age of the universe at the source redshift.  We allowed several
different SFH scenarios, including a single
burst, continuous star formation, and both the exponentially decaying and rising
(so-called ``tau'' and ``inverted-tau'') models.
We included nebular emission lines using the
prescription of \citet{salmon15}, which takes the line ratios
from \citet{inoue11}, assuming that the gas has the same metallicity
as the stars and that all the ionizing photons emitted by the model stellar
population are reprocessed in the galaxy and their escape is negligible.
To the rest frame spectra we added dust attenuation using the starburst attenuation
curve of \citet{calzetti00} in the range of 0 $\leq E(B-V) \leq$
0.8 ($0 \leq A_V\leq 3.2$ mag).
Then we redshifted the models to $0 < z < 11$
and added intergalactic medium (IGM) attenuation \citep{madau95}.
The resulting model spectra were integrated through our {\it HST} and
{\it Spitzer} filter
bandpasses to derive synthetic photometry for comparison with our
observations.  

We emphasize that  our model parameterization assumes that  the SFH of
each object follows  one of the simple scenarios  (i.e., single burst,
continuous,  tau,  or  inverted-tau),  not  a  superposition  of  such
scenarios. This  may seem like an oversimplification,  but evidence is
mounting that  the SFHs  in distant galaxies  vary smoothly  with time
\citep[e.g.][]{papovich11,finlator11,reddy12b,salmon15}.     Therefore,
the simple  scenarios may  in fact be  rather good  approximations, in
particular  when  deriving  the stellar  mass  \citep[e.g.,][]{lee10}.
Therefore, the  SFHs and  thus the physical  properties of  the bright
galaxies  that we  consider here  should not  be strongly  affected by
burstiness \citep[e.g.,][]{jaacks12b}.

The best-fit model was found via $\chi^2$ minimization.  We included an
extra systematic error of 5\% of the object flux in each band to
crudely account for the residual uncertainties in the zero point
correction and PSF matching process.  The uncertainties in the
best-fit parameters were derived via Monte Carlo simulations,
perturbing the observed flux of each object in each filter with a number drawn from a Gaussian 
distribution with a standard deviation equal to the flux uncertainty in the filter. 
Taking the source redshift to be statistically uncorrelated with other
spectral energy distribution (SED) fitting parameters, 
in each Monte-Carlo realization we drew the redshift from the photometric redshift statistical likelihood
function of the given object.  To prevent low-redshift solutions from
biasing the physical parameters, we limited the random redshift to be
within $\Delta z = \pm1$  of the best-fit
photometric redshift. This treatment of the source redshift effectively folded the uncertainty in redshift into the
uncertainty in the physical parameters \citep[most notably, the stellar mass
and $M_{1500}$;][]{finkelstein12b}.   For each galaxy, 10$^3$ Monte Carlo realizations were
generated and this provided a sample of as many values for
each model parameter.  SFRs for the best-fit model and for each Monte
Carlo realization,
were derived by converting the dust-corrected value of $M_{1500}$ to a
SFR via the relation of \citet{kennicutt98}.

For our subsequent analysis, we discarded realizations with poor best-fit
models with $\chi^2 >$ 20 to ensure robustness of the derived
properties.  This removed a relatively small number of galaxies (16,
9, 1, 0 and 1 at $z =$ 4, 5, 6, 7 and 8, respectively).  
The final sample contains 94, 46, 19, 14, and 1 galaxy at $z =$ 4, 5, 6, 7, and 8,
respectively.  As there is only one galaxy at $z =8$ that satisfies
these criteria, we focus on 4 $< z <$ 7 for the remainder of this
paper. Figure~\ref{fig:meansed}
shows the SED fit for a typical galaxy in each of our redshift bins.

\begin{figure}[!t]
\epsscale{1.1}
\plotone{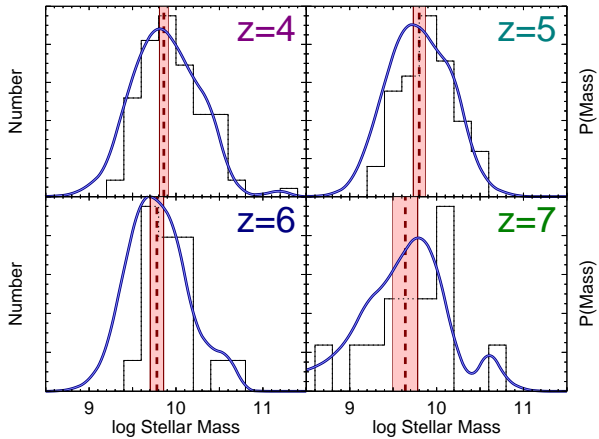}
\caption{Joint probability distributions for the stellar masses of bright
  galaxies ($M_{1500} < -$21) in our $z =$ 4, 5, 6 and 7 galaxy samples
  (blue line).  The histograms are
of the best-fit values, and the red dashed line and red shaded regions
denote the median and 1$\sigma$ uncertainty on the median, respectively, of the
joint PDFs.  Little significant evolution in the median stellar mass is seen over
this redshift range.}
\label{fig:mass}
\end{figure}

\subsection{Physical Properties}

The results of SED fitting are summarized in Table~\ref{tab:stellarpops}.
To derive the median stellar population properties, rather than stacking
the images or fluxes of the galaxies, we stacked samples of Monte Carlo realizations  
 in the model parameter space.  In each redshift bin, the stacked sample allowed us 
 to quantify the multivariate distribution of galaxy properties. The joint probability distribution functions for stellar mass marginalized over other parameters are shown
in Figure~\ref{fig:mass}.
 The median of a parameter such as the stellar mass is 
 taken to be the median value in the bin.
The $1\sigma$ confidence interval on the median was calculated via
10$^3$ bootstrap simulations where we rederived the median from a
randomly drawn (with replacement) sample of the galaxies in each
redshift bin.

As shown in Table~\ref{tab:stellarpops}, the median stellar population
parameters do not evolve significantly with redshift.
Broadly speaking, $M_\mathrm{UV}<-21$ galaxies at $z =$
4--7 are moderately massive ($\log[M_\ast/M_\odot] \approx$ 9.6--9.9), somewhat young ($<$100
Myr), and have non-negligible dust attenuation ($E(B-V) =$ 0.07--0.13
corresponding to $A_{V} \approx$
0.3--0.5 mag) and high SFRs ($\sim$40--60 $M_\odot$\ yr$^{-1}$).  

\begin{deluxetable}{cccccc}
\tabletypesize{\small}
\tablecaption{Median Physical Properties of Galaxies with $M_{1500} < -$21}
\tablewidth{0pt}
\tablehead{
\colhead{Redshift} & \colhead{Number} & \colhead{$\log(M_\ast/M_\odot)$} & \colhead{Age} &
\colhead{$E(B-V)$}  &
\colhead{SFR} \\
\colhead{$ $} & \colhead{$ $} & \colhead{$ $}  &
\colhead{(Myr)}  & \colhead{$ $}  &
\colhead{($M$\sol\ yr$^{-1}$)}  
}
\startdata
z = 4&94&9.86 $\pm$ 0.04&44 $\pm$ 2&0.13 $\pm$ 0.01&56 $\pm$ \phantom{1}4\\
z = 5&46&9.80 $\pm$ 0.06&35 $\pm$ 2&0.12 $\pm$ 0.02&52 $\pm$ 10\\
z = 6&19&9.78 $\pm$ 0.07&40 $\pm$ 4&0.07 $\pm$ 0.02&40 $\pm$ \phantom{1}8\\
z = 7&14&9.64 $\pm$ 0.13&29 $\pm$ 8&0.09 $\pm$ 0.02&41 $\pm$ 9 
\enddata
\tablecomments{Median values of physical parameters from the
  joint probability distribution describing all
  galaxies in a given redshift bin that have a
  measurement in the IRAC 3.6 $\mu$m channel and a best-fit model
  with $\chi^2 < 20$.  The number of galaxies satisfying these
  criteria in each redshift bin is given in the second column. The statistical uncertainties on median values
  were derived via $10^3$ Monte-Carlo simulations in which the median was rederived 
  from a randomly drawn sample (with replacement) of the galaxies in
  each redshift bin.  The spread in values for individual galaxies is larger.}
\label{tab:stellarpops}
\end{deluxetable}

Significant amounts of dust are likely produced in galaxies as early
as $z \sim7$, as \citet{finkelstein12a} and \citet{bouwens14} have
previously noted that massive and/or UV-bright galaxies had similarly
red UV continuum slopes at $z =$ 4--7, with 
a typical value of the UV spectral slope $\beta\sim-1.8$.  Both studies concluded that this
implied a similar amount of dust in bright/massive $z =$ 4--7
galaxies, independent of redshift.  We confirm this result,
finding $E(B-V)$ $\sim$ 0.1 in bright galaxies at $z
=$ 4--7 (constrained at 68\% confidence to be $>$ 0 and
$\lesssim 0.15$).   Therefore, although fainter/lower-mass galaxies
appear to be less dusty at higher redshifts
\citep[e.g.,][]{finkelstein12a,bouwens14}, this is not true for the
brightest galaxies.  This can be confirmed
with ALMA, and in fact ALMA has recently detected dust emission from
normal galaxies out to $z \sim$ 5--7.5 \citep[e.g.,][]{watson15,capak15}.

If the amount of UV attenuation due to dust among bright galaxies had evolved from $z
=7$ to 4, then it could have led to our selecting a lower stellar mass at
a given UV absolute magnitude (a similar effect could occur if the
dust attenuation curve is redshift dependent).  However, our results show that this is
not the case, as not only does our inferred attenuation exhibit no
evolution, but the median stellar mass also appears roughly
constant over the redshift interval.  

Our fiducial analysis assumed a dust attenuation curve from
\citet{calzetti00} for consistency with previous results in the
literature.  However, a number of recent studies have found that a
Small Magellanic Cloud (SMC)-like \citep{pei92} attenuation curve may be more appropriate for
high-redshift galaxies \citep[e.g.,][]{reddy12,tilvi13,capak15}.  
This dust attenuation curve has more attenuation for given values of
$E(B-V)$, thus we explored how our results would change had we assumed
this attenuation curve in our fiducial analysis.  Even with this
different attenuation curve, the amount of
dust attenuation appears to be roughly constant in these bright
galaxies, with slightly lower values of E(B-V) $=$ 0.06 -- 0.08. 
We find that our median stellar masses would
change by at most 0.14 dex (consistent with results from
\citealt{papovich01} that the choice of attenuation curve does not
significantly affect stellar mass measurements), resulting in a minimal change to our major
conclusion below on the stellar baryon fraction (although the slope of the
stellar baryon fraction evolution assuming SMC dust is less than our
fiducial scenario, and thus evolution is detected at reduced
significance due to a lower value and larger
uncertainty on the median mass of the z=7 sample in this scenario). 
We conclude that our assumption of a Calzetti et al.\ attenuation curve does not result in
a strong bias in our results.

The youngish character of these galaxies is surprising.  Although age is notoriously difficult to
measure robustly, the addition of the IRAC photometry does help,
in particular at $z =$ 6, where the ensemble is constrained to have a typical
age $<$ 50 Myr (68\% C.L.).
While  evolved galaxies are not absent at these high
redshifts (for example, z7\_GNW\_17001 has $\log(M_\ast/M_\odot) = 10.7$ and an
age of 400 Myr), they seem to be exceptions. This is in stark contrast with galaxies at $z \lesssim 3$ , where
those residing at the bright-end of the luminosity function tend to be
more evolved.   

\citet{reddy06} published a stellar population analysis
of galaxies at redshifts $z \sim$ 1--3.5.  Using the results from their
models that assume a constant SFH, the median galaxy
with $M_\mathrm{UV} < -21$ (derived from  the observed R-band magnitude and
spectroscopic redshift) at $z \sim$ 2 -- 3 has $\log(M_\ast/M_\odot) = 10.2$, an age of 260 Myr, $E(B-V)
=0.16$ and SFR $=$ 70 $M_\odot$\ yr$^{-1}$ (the results are similar when the sample is split into two redshift bins centered at
$z =$ 2 and $z =$ 3).  Thus, bright galaxies at the peak of cosmic
star-formation activity have similar SFRs and dust attenuations as bright
galaxies in the epoch of reionization, but the lower redshift galaxies
are $\sim$ 2--2.5$\times$ more massive and have rest-UV/optical light
dominated by stars $>$5$\times$ older.
Such a comparison with lower redshift makes sense if one is interested in the evolution of
properties of similarly bright galaxies with redshift.  It is also
valid if we wish to explore the evolution of galaxies with $M_\mathrm{UV} <
M_\mathrm{UV}^{\ast}$ with redshift, as $M_\mathrm{UV}^{\ast}$ at $z \sim 2.3$ and 3
\citep[$-20.7$ and $-20.97$, respectively;][]{reddy09} are similar to what we
find at $z \geq 4$.  

A more interesting question is how UV-bright
galaxies at $z \sim$ 2--3 are related to UV-bright galaxies at $z
\sim$ 6--8.  Specifically, in view of the hierarchical galaxy
assembly, are the former galaxies descendants of the latter?  Are the
latter progenitors of the former?  
Moreover, galaxy merging complicates direct number-counting-based matching across redshifts.  
A few recent studies have tried  
comparing galaxies at different redshifts at the same cumulative
number density \citep[e.g.,][]{vandokkum10,papovich11,leja13}.  
\citet{behroozi13c} showed that such a comparison is adequate for
identifying the low-$z$ descendant population of a high-$z$
population \citep[see also][]{jaacks15}. This is because the majority of massive high-$z$ galaxies do
not end up merging into substantially more massive systems.
However, the converse is not true: reflecting hierarchal merging, the cumulative number density of
the high-$z$ progenitor population of a low-$z$ population has a
comparatively higher cumulative number density.

As shown in the following subsection, galaxies at $z
=$ 6 and 7 have $n(M_{\rm UV}< - 21) = 3.6 \times 10^{-5}$ Mpc$^{-3}$ and
2.4 $\times$ 10$^{-5}$ Mpc$^{-3}$,
respectively.  Using the luminosity functions of \citet{reddy09} at
 1.9 $< z <$ 2.7 (the median redshift of the $z \sim$ 2--3
comparison sample we quote here is $z =$ 2.44), we find $n(M_{\rm UV}< - 21)
= 2.24 \times 10^{-4}$ Mpc$^{-3}$, unsurprisingly more abundant
than our high-redshift sample.  To match the cumulative number
densities of our high-redshift samples, we need to select objects with $M_{\rm UV} <-20.0$ and $< -19.8$ for our $z =6$ and 7 sample, respectively.
The $M_{\rm UV}<-21$ galaxies at $z \sim$ 2--3 are thus plausible descendants of the high-$z$ galaxies at these fainter magnitudes.

Although above we
only report stellar population results for bright galaxies, we
performed SED fitting on the entire sample of
\citet{finkelstein15}, finding that $M_\mathrm{UV}
= -$20 galaxies at $z =$ 6 have mean stellar masses of $\log(M_\ast/M_\odot)
= 8.7$ and mean ages of 34 Myr.  At $z =$ 7, $M_\mathrm{UV}
= -19.8$ galaxies have median stellar masses of $\log(M_\ast/M_\odot)
= 8.5$ and median ages of 28 Myr.  We conclude that, compared to either similarly bright
or similarly abundant galaxies at lower-redshift, UV-bright galaxies
at $z >$ 6 are significantly younger and less massive, but have
similar SFRs, as inferred from their similar UV luminosities, and exhibit
relatively little evolution of dust attenuation at these luminosities.

We acknowledge that ages are notoriously difficult to
constrain, as they are tied to the assumed
star-formation histories \citep[e.g.,][]{papovich01}.  In more realistic scenarios, the young stars
that dominate the observed SED are modestly "outshining" the older
generations, potentially biasing the measured age.  However,
the stellar masses (which we use for our primary result in the
following sections) are robust to age variations \citep{lee10}, thus
this potential bias does not affect our main results.

\begin{figure*}[!t]
\epsscale{0.65}
\plotone{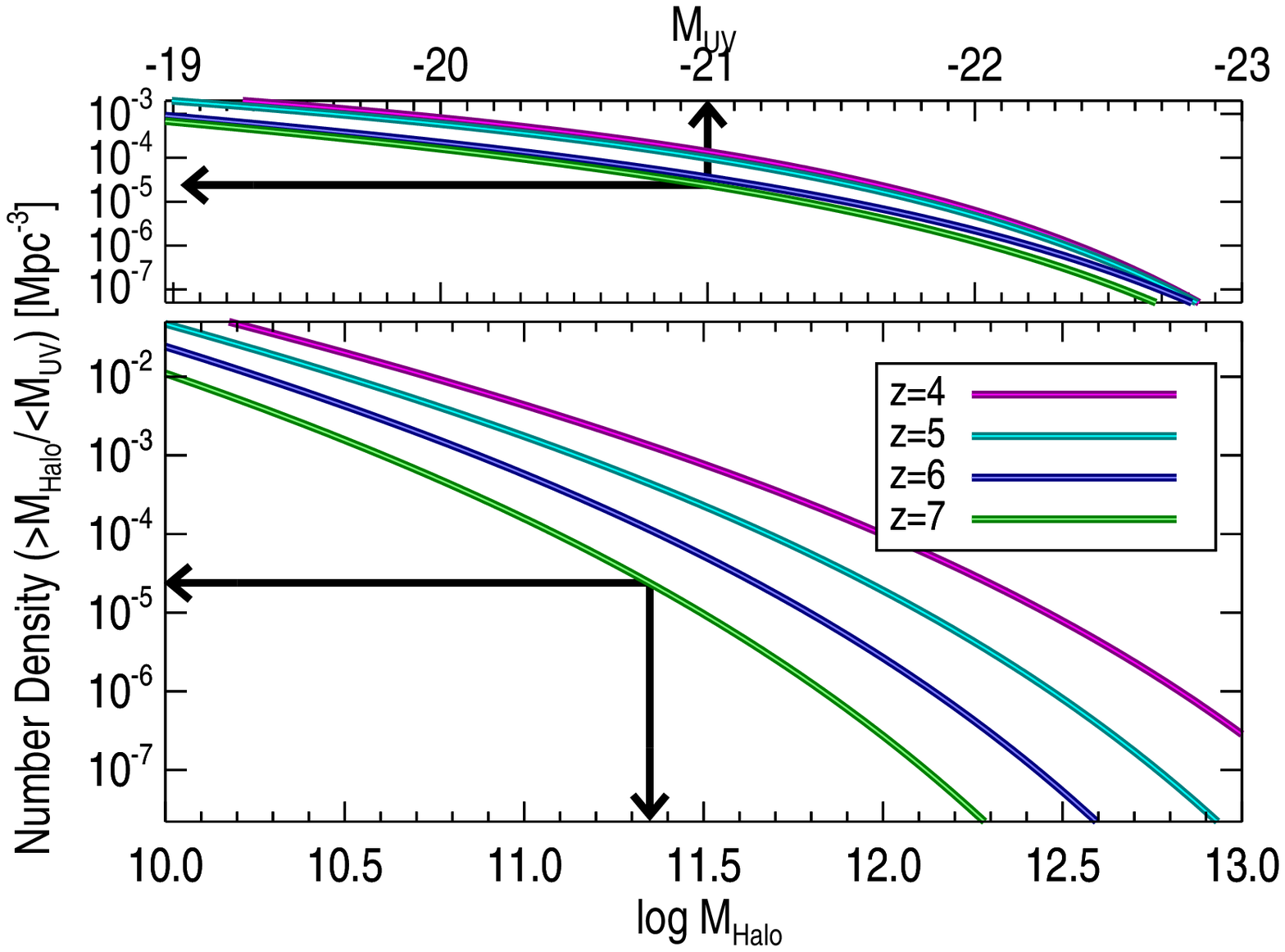}
\hspace{2mm}
\raisebox{0.13\height}{\includegraphics[height=2.15in]{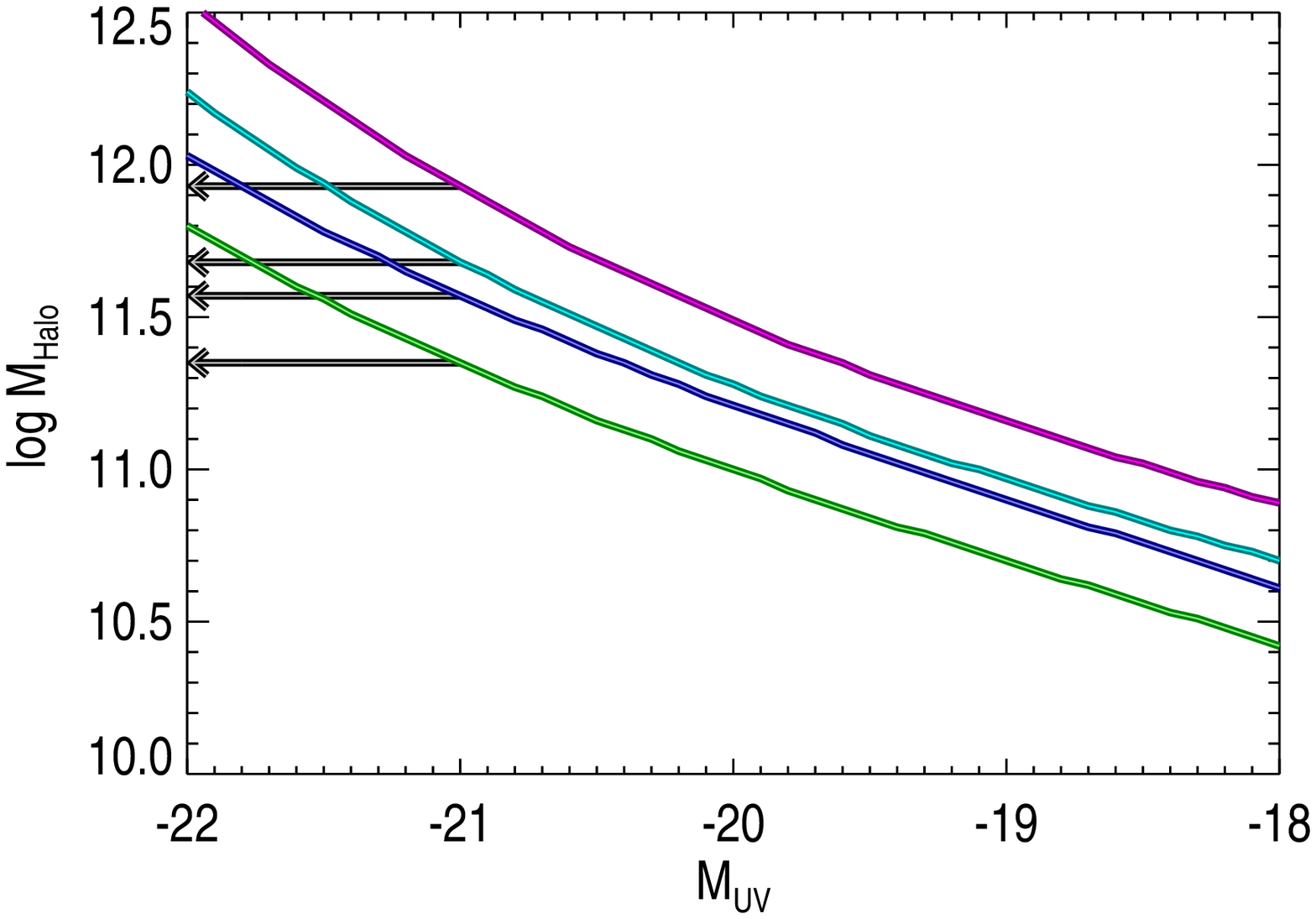}}
\caption{Top left: The cumulative luminosity function at $z
  =$ 4, 5, 6, and 7.  Bottom Left: Cumulative halo mass functions
  at $z =$ 4, 5, 6, and 7, derived by volume-averaging the Bolshoi snapshot mass functions
over the same redshift ranges as those defining our galaxy samples.
The arrows show our results from abundance matching at
  $z = 7$, where galaxies with $M_{1500} < -21$, which have $n(M_{\rm UV} <-21) =
  2.5 \times10^{-5}$ Mpc$^3$, have halo masses of
 $\log(M_{\rm h}/M_\odot) = 11.35$.  Right: Relation between observed UV
  absolute magnitude and abundance-matching-derived halo mass at our
  redshifts of interest.  The arrows denote the halo masses at our
  magnitude of interest of $M_{\rm UV}=-21$.}
\label{fig:cmf}
\end{figure*}

\section{Halo Masses of UV-Bright Galaxies}

\subsection{Halo Masses}
\label{sec:halo_masses}

Given the relatively high stellar masses and SFRs seen in our sample of $z \geq6$  galaxies, one may wonder if the processes governing
gas cooling and the conversion of cold gas into stars
differ from those at lower redshift.  As a step toward answering this question, 
we compare the
stellar masses $M_{\ast}$ of these systems to the baryon masses in
halos ($\Omega_{\rm b}/\Omega_{\rm m})\,M_{\rm h}$ computed assuming
a cosmic-mean baryon fraction of $\Omega_{\rm b}/\Omega_{\rm m} =$
0.1669 \citep{komatsu11}.
The clustering of these systems would have provided the most direct
constraints on the halo mass, but the numbers and surface densities of
these galaxies, particularly at $z \geq7$, are not yet sufficient to
permit a robust clustering analysis \citep[though see][]{baronenugent14}.  Instead, we use abundance matching 
to estimate the halo masses for our bright
galaxies.  We refer the reader to previous works for
a full discussion of abundance matching
\citep[e.g.,][]{moster10,behroozi10}.  Here, we review the procedure
only briefly. 

Abundance matching assumes that the galaxy luminosity or stellar mass is a monotonic function of the halo mass. The most
luminous galaxies are assumed to live in the most massive halos.  This is certainly a plausible assumption among the luminous galaxies we study here, but perhaps not among the more stochastic dwarf galaxies.  Starting with a 
cosmological simulation (in this case, the pure $\Lambda$CDM simulation Bolshoi), one selects simulation snapshots close in redshift to the target redshift of an observational survey.  In these snapshots,
one selects a simulation volume equal to the 
volume of the survey.  One then identifies all virialized halos in the volume and rank-orders them by mass.
After rank-ordering the observed galaxies by their luminosities, one places
each galaxy in the simulated halo of the matching rank.  
This procedure provides a mapping of galaxy statistics onto host halo statistics.  

Here, we use Schechter parameterization of the
observed luminosity functions at each redshift from \citet{finkelstein15} to estimate the
host halo masses for the bright ($M_{1500} < -21$) galaxies in
our sample.  Given the present stellar mass function uncertainties, luminosity-based matching is currently more robust than the
stellar mass-based matching.  
In the future, the matching should be performed directly with the
stellar mass as it should be more correlated than the UV luminosity with the host halo
mass \citep[e.g.,][]{lee09,gerke13}.

For our analysis, we use the results from the Bolshoi cosmological
simulations \citep{klypin11}, which has 2048$^3$ particles in a 250
(Mpc$/h)^3$ box. This translates to a halo mass
resolution of $\log(M_{\rm h}/M_\odot) =10$.  
As we will find that our galaxies have halo masses $\gg10^{10}$ M\sol,
the resolution of Bolshoi is more than sufficient for the present analysis.
We use the halo mass functions derived in \citet{behroozi13b}, which
are a modification of the \citet{tinker08} mass
functions, include subhalos, and are accurate at very high redshifts \citep[see Appendix G of][]{behroozi13b}.
We derived halo mass functions at our redshifts
of interest by volume-averaging the Bolshoi snapshot mass functions
over the same redshift ranges as those defining our galaxy samples. 
 
The top-left panel of Figure~\ref{fig:cmf} shows the cumulative luminosity
functions at $z=$ 4--7 and the bottom-left panel shows the cumulative halo mass
functions at the same redshifts.  
To infer the host halo masses for the bright galaxies with $M_{1500} < -21$, we first find
the cumulative number density of galaxies at that magnitude. For example, for $z=7$, the number density is
indicated with the horizontal arrow in the two left panels of
Figure~\ref{fig:cmf}.  The host halo masses can then be inferred by
finding the halo mass above which the cumulative halo number density
equals the cumulative galaxy density.  We find $\log(M_{\rm
  h}/M_\odot) = 11.93$, 11.68, 11.57, and 11.35 at $z =4$,
5, 6, and 7, respectively.  To estimate the uncertainties in the halo
masses, we ran $10^3$ Monte Carlo simulations, in each drawing a
luminosity function randomly from the MCMC
sample generated during the luminosity function estimation, which account for both Poisson noise
and uncertainties in the luminosity function completeness simulations \citep[see][]{finkelstein15}.  We find that the uncertainties in
the halo mass are low, $\sim$0.01--0.03 dex, reflecting relatively low uncertainties in the
cumulative luminosity functions.

\subsubsection{Cosmic Variance}

To estimate how our measured abundances of bright galaxies
are affected by cosmic variance, we used a suite of semi-analytic
models.  These models, based on \citet{somerville08}, were provided to
the CANDELS team, and include a set of mock catalogs, one for each of the five CANDELS fields (with each individual mock catalog
covering a volume somewhat larger than the observed volume).  We
tuned the dust attenuation in these SAMs to match
the observed UV luminosity functions from \citet{finkelstein15}.  In each of our redshift bins, we
used the SAMS to extract 64 independent volume elements comparable to
one GOODS-sized field ($16^\prime \times10^\prime$) and 3192
independent volume elements comparable to a single WFC3 pointing
($2.1^\prime \times 2.1^\prime$).  In each redshift bin, the fractional uncertainty due to
the combination of Poisson fluctuations and cosmic variance was derived as the standard deviation (computed over all of the realizations of a given field size) 
of the number of
galaxies in a luminosity bin of $\Delta M_\mathrm{UV}=0.5$ mag centered at
$M_\mathrm{UV}=-21$, divided by the mean number of galaxies in the
bin.  

The total fractional uncertainty due to Poisson fluctuations and cosmic variance was then derived by
combining the variances from two GOODS-sized fields and
five single WFC3 pointing-sized fields, as this was the area used in
the luminosity function calculation by \citet{finkelstein15}, who used
GOODS-S, GOODS-N, and five individual fields (the HUDF, the two HUDF
parallels, and the first two first-year Frontier Fields parallel fields).  We
find that the fractional uncertainties due to cosmic variance are 0.132,
0.159, 0.212, and 0.327 at $z =$ 4, 5, 6, and 7, respectively.  We
include these uncertainties in the Monte Carlo simulations discussed
in the above paragraph, and find that the uncertainty in the derived
halo mass increases by a factor of $\sim2$ to 0.03 dex at $z =$ 4 and
0.06 dex at $z =$ 7 with the inclusion of the uncertainty due to
cosmic variance.  Our derived halo masses, and these total
uncertainties, are listed in Table 2.

\subsection{Evolution of the Stellar-to-Halo Mass Ratio}

Comparing the halo masses estimated in the previous
subsection to the median stellar masses estimated in \S 2.3, we can calculate the ratio of the median
stellar mass to halo mass (SMHM).  We find $M_{\ast}/M_{\rm h}
= 0.009$, 0.013, 0.016, and 0.020 at $z = 4$, 5, 6, and 7,
respectively.  Thus, at a constant UV luminosity,
the stellar-to-halo mass ratio increases with increasing redshift.  These results
are listed in Table 2.

\begin{figure*}[!t]
\epsscale{0.85}
\plotone{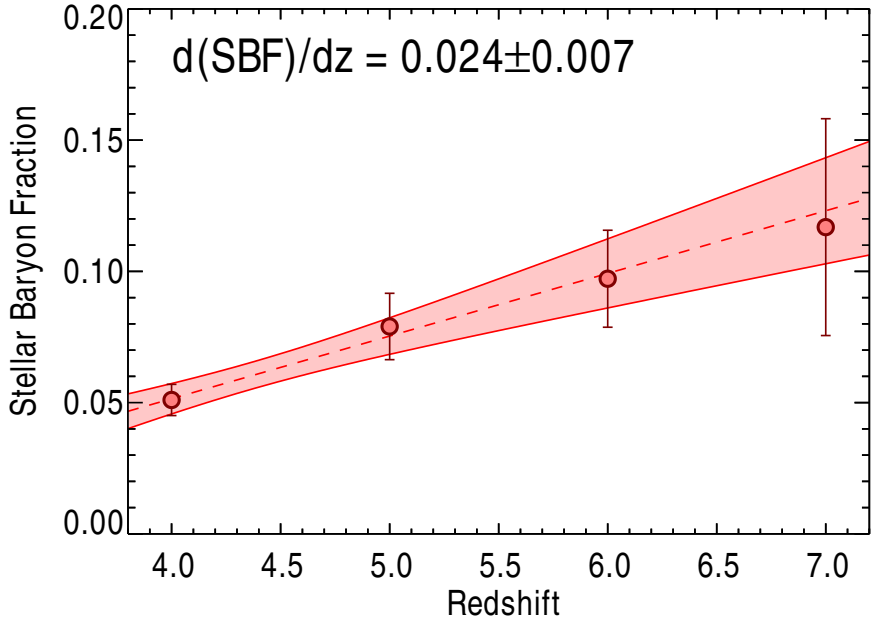}
\caption{The stellar baryon fraction (SBF) in bright ($M_\mathrm{UV} = -21$)
  galaxies from $z =4$ to 7.  We define the SBF as the
  stellar to halo mass ratio in units of the cosmic baryon mass fraction $\Omega_{\rm b}/\Omega_{\rm m}$.  
  We find that the SBF increases with
  increasing redshift, which may be responsible for the apparent lack
  of evolution in the characteristic magnitude $M_\mathrm{UV}^{\ast}$
  observed over this redshift range.}
\label{fig:sfe}
\end{figure*}

The cosmic baryon mass fraction is much higher than the
$M_{\ast}/M_{\rm h}$ ratios we derive, at
$\Omega_{\rm b}/\Omega_{\rm m} =0.1669$
\citep{komatsu11}.  In Figure~\ref{fig:sfe}, we show the evolution of
the SMHM ratio in the units of 
$\Omega_{\rm b}/\Omega_{\rm m}$ as a function of redshift.  We refer to this quantity as the stellar
baryon fraction (SBF), as it measures the amount of baryons converted into
stars compared to the cosmic allotment of baryons in the halo.  We
find that the SBF 
evolves from  $0.117 \pm 0.043$ at $z =
7$ to $0.051 \pm 0.006$ at $z =4$.  This factor of $\sim3$ evolution is significant, as fitting a linear function for $\mathrm{SBF}(z)$ yields the slope $d\mathrm{SBF}/dz
= 0.0239 \pm 0.0074$, with the trend detected at $3.2\sigma$ significance.  

The inferred significance of the trend depends on our assumed uncertainties in the
stellar baryon fractions.  We assumed an uncertainty in the halo mass
and median stellar mass as reported in Table
2.  The quoted stellar mass uncertainties, which decrease from 0.14 dex at $z =7$ to 0.04 dex at
$z =4$, are well below the
uncertainty for an individual object, as here we are interested in the
accuracy with which we can constrain the median stellar mass at
different redshifts.  However, these values are
consistent with the typical uncertainty in the
median stellar mass at $M_{\rm UV} = -21$ derived by \citet{song15}, who fit
the mass-to-light ratio over a wide dynamic range in UV luminosity, finding that it
decreases from 0.1 dex at $z =7$ to 0.02 dex at $z =4$.  
If one used the median of 
individual galaxy mass uncertainties in each of our
samples, which decreases from 0.19 dex at $z =7$ to 0.11 dex at $z =4$, 
the trend of increasing SBF with redshift is
still apparent (albeit reduced in significance to $\sim1.7\sigma$).

Finally, we recall that our fiducial sample was selected to include galaxies with
$M_\mathrm{UV} < -21$.  Given the shape of the luminosity function,
the median magnitude of this sample is 
very close to $-21$ (ranging from $-21.2$ to $-21.3$).  To check if our sample selection
biases the median masses, we examined how the evolution of
the stellar baryon fraction changes if we use the median stellar mass of galaxies with luminosities
$M_{\rm UV} = - 21 \pm 0.25$.  We find a comparable
number of galaxies as in our fiducial sample, specifically 118, 76, 20, and 17
galaxies at $z =4$, 5, 6, and 7, respectively.  The median stellar mass is slightly
lower than in our fiducial sample, $\log(M_\ast/M_\odot) = 9.68$, 9.66, 9.50 and
9.52 and $z =4$, 5, 6 and 7, respectively.  The amplitude of the
redshift derivative of the stellar baryon fraction is thus lower, 
$d\mathrm{SBF}/dz = 0.0123 \pm 0.0041$, yet the evolution is still
significant at the $3.1\sigma$ level.  We conclude that there is
significant evolution in the stellar baryon fraction in that it decreases with decreasing redshift, 
and that this evolution is stable against
several definitions of both the median stellar mass and the stellar mass uncertainty.

\subsubsection{Comparison to Previous Results}

\citet{behroozi13b} recently studied the evolution of the SMHM
relation by modeling all available observational constraints, including luminosity
functions, stellar mass functions, and SFRs, and 
exploring the galaxy evolution parameter space with an MCMC search.  They found that the SMHM
curve peaks at a roughly constant halo mass of  $\log(M_{\rm h}/M_\odot) = 11.7$
at $z =$ 0--4.  Specifically, at $z = 4$, they found a peak SBF at a
halo mass of $\log(M_{\rm h}/M_\odot) \approx 12.0$, consistent with the typical halo mass we
derive for our bright $z =4$ galaxies, $\log(M_{\rm h}/M_\odot) = 11.9$.
Then they went on to find that the peak of the SMHM
relation at $z \geq$ 5 occurs at a halo mass steadily decreasing with redshift,  
$\log(M_{\rm h}/M_\odot) = 11.9$, 11.6, and 11.4 at $z = 5$, 6, and 7,
respectively.  This is very similar to what we find for the typical
halo masses of $M^{\ast}_{\rm UV}$ galaxies, $\log(M_{\rm h}/M_\odot) = 11.7$, 11.6,
and 11.4 at $z =5$, 6, and 7, respectively.  
The peak SMHM ratio measured by \citet{behroozi13b} is
somewhat higher, $\sim$3.7\% at $z =7$ (converting from a Chabrier to
our Salpeter IMF) than the 2.0\% we find here.  
However the two values are indistinguishable 
given the significant
uncertainties in the high-redshift observables, particularly the stellar mass functions used by
\citet{behroozi13b}.

Although here we are specifically concerned with the halo masses of
bright galaxies, in the right-hand panel of
Figure~\ref{fig:cmf} we also provide the abundance-matching-derived halo masses at all observed
magnitudes in each of our redshift bins.  
Our derived halo mass of $\log(M_{\rm h}/M_\odot) = 11.35$ at $z =7$
is consistent with a recent clustering-based measure of $\log(M_{\rm h}/M_\odot)
\approx 11.2$ from \citet{baronenugent14}.  However, their halo mass estimate was for
galaxies with $M_{\rm UV} < -19.4$, a sample that has a lower average
luminosity than our sample, and thus a lower halo mass is expected, as
shown in the right-hand panel of Figure 3.  Our $z = 4$
results are consistent with earlier clustering-based estimates 
by \citet{lee06}, while at $z =$ 5 and $z =$ 6, our derived halo
masses are somewhat higher than the clustering-based estimates from
\citet{lee06} and \citet{overzier06}, respectively,  likely due to the fainter
luminosities considered in those works ($M_\mathrm{UV} < - 20$
$-$19.5, respectively). Although there are minor
differences, it is encouraging that two
independent methods, abundance matching and clustering, tentatively
agree on the halo mass estimates at such high redshifts, though
certainly the clustering-based can be made more robust at
$z >$ 6.

\begin{deluxetable}{ccccc}
\tabletypesize{\small}
\tablecaption{Dark Matter Halo Properties of Bright Galaxies}
\tablewidth{0pt}
\tablehead{
\colhead{Redshift} & \colhead{$\log n (M_{\rm UV}<-21)$} & \colhead{$\log
  M_\mathrm{h}$} & \colhead{Median} & \colhead{Stellar Baryon} \\
\colhead{$ $} & \colhead{(Mpc$^{-3}$)} & \colhead{($M$\sol)} &
\colhead{$M_{\ast}/M_\mathrm{h}$} & \colhead{Fraction}
}
\startdata
z = 4&$-$3.86&11.93$^{+0.03}_{-0.03}$&0.009$\pm$0.001&0.051$\pm$0.006\\
z = 5&$-$4.01&11.68$^{+0.04}_{-0.03}$&0.013$\pm$0.002&0.079$\pm$0.013\\
z = 6&$-$4.45&11.57$^{+0.06}_{-0.03}$&0.016$\pm$0.003&0.097$\pm$0.019\\
z = 7&$-$4.62&11.35$^{+0.09}_{-0.06}$&0.020$\pm$0.007&0.117$\pm$0.043\\
\enddata
\tablecomments{The uncertainties in the halo mass are derived via
  Monte Carlo simulations and include the uncertainty in the number
  density of $M_{\rm UV} < -21$ galaxies, which reflects our fiducial luminosity
  function uncertainties as well as cosmic
  variance.  The uncertainties in $M_{\ast}/M_\mathrm{h}$ and 
the stellar baryon fraction assume an uncertainty in the median stellar
mass from Table 1.}
\label{tab:halo}
\end{deluxetable}

\subsubsection{UV Luminosity Scatter}

The halo mass estimates in \S~\ref{sec:halo_masses} did not explicitly
model the effect of the scatter in UV luminosity at a fixed halo
mass.  The estimates compared the number density in galaxies with $M_{\rm UV}<-21$ to the number density in halos with masses $>M_{\rm h}$ to derive the characteristic halo masses $M_{\rm h}$.  The halo mass-luminosity relation has an intrinsic scatter that is usually treated as a Gaussian in the magnitude $M_{\rm UV}$ centered on some median (or mean) halo-mass-dependent magnitude.  Because of this scatter, some halos with masses $>M_{\rm h}$ can can host galaxies with atypically low luminosities, $M_{\rm UV}>-21$, placing them outside of our luminosity cut. However, in \S~\ref{sec:halo_masses}, the halos hosting these faint interlopers were being counted in the cumulative halo mass function and this could have biased the derived $M_{\rm h}$.  If we had known which halos hosted faint interlopers (which we do not), we would have excluded them from the halo counting to derive a more accurate $M_{\rm h}$. Excluding a fraction of halos at every halo mass, the resulting characteristic halo mass $M_{\rm h}$ would have been lower than the scatter-blind estimate  \citep{behroozi10}.

Here, we attempt to quantify this bias.
We estimate the amplitude of the
luminosity scatter from the relation between the UV
luminosity and the stellar mass derived by \citet{song15}. This is
warranted because the stellar mass is expected to
be more correlated with the halo mass than with UV luminosity.
We find that at the mass of our sample $\log (M_\ast/M_\odot)
\sim 9.8$, the scatter in the UV absolute magnitude is $\sim0.3$ dex
at all redshifts we consider.

To assess the impact of luminosity scatter on our abundance
matching-determined halo masses, we carried out the iterative deconvolution described in \citet{reddick13} to get a handle on the true, underlying bivariate distribution of halo masses and observed UV luminosities that exhibits an intrinsic luminosity scatter at fixed halo mass.  We performed abundance matching on the bivariate distribution properly applying the $M_{\rm UV}<-21$ cut to find halo masses $\log(M_{\rm h}/M_\odot) = 11.65$, 11.44, 11.36, and 11.13 at $z
= 4$, 5, 6, and 7, respectively.  These values are $\sim0.2$ -- $0.25$ dex
lower than the values obtained by scatter-blind abundance matching in \S~\ref{sec:halo_masses}.  Therefore we expect that the true host halo masses of our $M_{\rm UV}<-21$ galaxies are slightly lower than those reported in Table 2.
If the luminosity scatter had been 0.5 dex, our
halo masses would be overestimated by 0.7--0.9 dex.

Since halo masses at all redshifts are corrected by a similar logarithmic increment, 
the effect of luminosity scatter does not
affect our detection of an evolving stellar baryon fraction.  But for
higher luminosity scatter which could characterize a different regime
in which galaxy formation is more stochastic, this effect could be
more significant.

\subsubsection{Systematic Uncertainties in Stellar Mass}

Our derivation of the stellar masses required a variety of
assumptions that could have potentially biased our results. Could any of the biases explain away
our conclusion that UV-bright galaxies from $z =4$  to 7 have
similar stellar masses but are found in progressively lower mass
halos toward higher redshifts?   One strong assumption we made is that of the Salpeter
IMF.  If the high-mass slope of the IMF evolved with redshift, this certainly could have
biased the conclusion.  Other assumptions involve the parameterization of SFHs and the
treatment of metallicities and dust attenuations.  
As discussed earlier, the UV-continuum slopes in UV-bright galaxies
appear to be roughly constant across this redshift range, and so dust
abundances will only affect our results if the typical dust-law
changes as a function of redshift.
Direct measurement of the metallicities is beyond our current
capabilities at these high redshifts \citep[though
see][]{finkelstein13}, but a plausible metallicity variation produces
only a minor change in colors.  The SFH parametrization  is
potentially more troubling, though recent results indicate
that at least on average, galaxy-scale SFRs are smooth functions of time
\citep[e.g.,][]{salmon15}.

\begin{figure}[t]
\epsscale{1.22}
\plotone{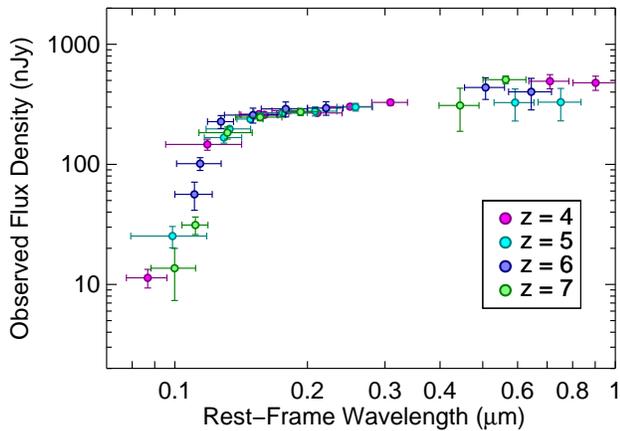}
\caption{The median flux-stacked SEDs of our samples of bright
  galaxies at each redshift.  The spectral shapes of these SEDs are
  remarkably similar, with minor differences appearing at longer
  wavelengths due to the presence of nebular emission lines, as well
  as the generally lower signal-to-noise of the long-wavelength data.
The near identical nature of these SEDs confirms our conclusion that
bright galaxies from $z =$ 4 to 7 are physically very similar, and that
these $\log(M_\ast/M_\odot) =$ 9.6--9.9 galaxies on average inhabit lower
mass halos at higher redshifts.}
\label{fig:stack}
\end{figure}

A basic test of any bias in the mass measurements is to compare the shapes of
the SEDs of our galaxies.  Figure~\ref{fig:stack} shows a median flux
stack of galaxies in each of our redshift bins versus the rest-frame
wavelength, scaled vertically to a common redshift $z=6$.  The shapes of the
SEDs are remarkably similar, especially in the rest-frame UV, where
they appear identical.  Modest differences are visible in the rest-frame
optical, likely due to the lower signal-to-noise of {\it
  Spitzer}/IRAC data, as well as the strong nebular [\ion{O}{3}] and
H$\alpha$ lines which redshift through the bandpasses.  
Given the highly similar SED shapes, it is unlikely that an
unaccounted for systematic effects strongly bias our results.
Rather, we appear to be studying a very similar type of galaxy at each
redshift; this type of galaxy lives in lower mass halos at higher
redshift.  This conclusion is confirmed by stellar population model fits to
the stacks, which yield stellar masses consistent within $\sim$0.3
dex of the median stellar masses in Table 1 ($\log [M_\ast/M_\odot] =$ 9.8,
10.0, 9.9 and 9.9 at $z
=4$, 5, 6, and 7, respectively).

\subsubsection{Dusty Star-Forming Galaxies}
Our sample is selected on the basis of UV luminosity and therefore it is
prudent to examine what effect that may have on our results. 
In particular, a rest-frame UV selection may miss extremely dusty
galaxies which have their UV light attenuated below our
detection sensitivity.  Such systems, referred to as sub-millimeter galaxies
(SMGs; after their selection wavelength), have been well studied at
moderate redshifts.  These galaxies are
typically very massive (log M/M\sol\ $>$ 11; \citealt{casey14}).  If
we were missing a large population of these galaxies, it would bias
our derived halo masses to be too high (as we would be placing our
observed galaxies in the most massive halos in our volume, which would
truly be occupied by these dusty galaxies).  If the
abundance of SMGs evolves with redshift at $z >$ 4, then not
accounting for these systems would bias our halo masses differently at
different redshifts, biasing our inferred evolution of the stellar
baryon fraction.

To explore the potential bias introduced by our UV-only selection, we
compare the space density of SMGs at $z =$ 4 to our UV luminosity
function.  The space density of SMGs at high redshift is very
uncertain for a variety of reasons, including the relatively shallow
depths of sub-millimeter surveys, and the difficulty of obtaining
redshifts for such systems.  We start at $z =$ 2, where
\citet{chapman05} found a space density for SMGs of 5 $\times$
10$^{-5}$ Mpc $^{-3}$.  A more complete survey for SMGs at multiple
wavelengths was performed by \citet{casey13}, who found a SMG SFR
density $\sim$2$\times$ higher (see Fig.\ 23 of \citealt{casey14}),
implying a space density of $z =$ 2 SMGs of 10$^{-4}$ Mpc $^{-3}$.  To
estimate the evolution in this quantity to $z =$ 4, we use the
redshift evolution shown by \citet{casey14}, which shows that the SFR
density from SMGs is an order of magnitude lower at $z \sim$ 4 than at
$z \sim$ 2.  Therefore we adopt 10$^{-5}$ Mpc $^{-3}$ as a fiducial space
density for $z =$ 4 SMGs.

We simulated the absence of SMGs in our sample by adding this abundance to
our observed cumulative luminosity functions, and repeating the
abundance matching analysis.  The absence of such a
population of SMGs results in a bias of the $z =$ 4 halo mass for our
galaxies of interest of only 0.02 dex (11.91 versus our fiducial
result of 11.93).  This is within the 68\% confidence range on our
fiducial halo mass, therefore not significant.  If the abundance of SMGs stays constant
to $z =$ 7, the declining UV luminosity function results in a slightly larger
overestimate of the halo mass of 0.06 dex at $z =$ 7.  Were this
the case, the observed evolution in the stellar baryon fraction would
be even stronger than we observe.  However, such a high abundance of
dusty galaxies at $z =$ 7 is highly unlikely.   We therefore conclude
that our selection does not affect our main
conclusions in this study.

\section{Discussion}
To understand the physical effects responsible for our
observed trend of an increasing SBF with increasing redshift, here we
consider a variety of possible mechanisms, with the caveat that our
observations cannot uniquely distinguish between these scenarios.
First, galactic gas at higher
redshifts has higher surface densities $\Sigma_{\rm g}\propto f_{\rm
  gas}\,M_{\rm h}^{1/3} (1+z)^2$, where $f_{\rm gas}$ is the gas
fraction in the cooled, virialized phase.  The typical free-fall time 
to which the gas-to-stars conversion rate is proportional 
\citep[albeit with a small coefficient---dimensionless SFR---see, e.g.,][]{krumholz12b} is
$t_{\rm FF} \propto (1+z)^{-3/2}$.  However,
masses of the most massive progenitors (MMP) of our galaxies' host
halos are $M_{\rm h,MMP}(z')=M_{\rm h}(z) e^{-\alpha(z'-z)}$ with
$\alpha\approx 1$, where $z'>z$ is the progenitor redshift
\citep{neistein06,fakhouri10}.  Therefore the mass-doubling growth
time is $t_{\rm grow}(z) \sim [(1+z) H(z) d\ln M_{\rm h,MMP}/dz]^{-1}
\propto (1+z)^{-5/2}$.  This means that the ratio of the free-fall
time to the growth time
\emph{increases} with increasing redshift as $t_{\rm FF}/t_{\rm
  grow}\propto 1+z$, suggesting that if the minor progenitors are
inefficient star formers so that they do not contribute substantial
stellar mass to the main branch, it should be progressively more
difficult at high redshifts to convert the gas acquired through growth
into stars \citep{bouche10,krumholz12,dekel13}.  
Second, the gas-phase metallicity and associated dust abundance appear to
decrease with increasing redshift and decreasing mass
\citep[e.g.,][]{finkelstein12a,bouwens14}.
Because metals and dust are
the principal gas-cooling and UV-shielding agents, the observed trend
of decreasing dust with increasing redshift
could have a dramatic effect on the abundance of the cold
($T\lesssim 1000\,\mathrm{K}$) gas in which star formation seems to
exclusively happen \citep{krumholz12,krumholz12b}.  The drop in cold gas abundance toward higher redshifts would imply lower SFRs.

All this suggests that from the supply-versus-consumption side alone, we
expect an opposite trend from the one measured, a trend in which the
stellar baryon fraction decreases with increasing redshift.  
How should we then interpret the stellar baryon fractions that increase
with increasing redshift?  We can only speculate.  Our
observations imply that cold gas is less readily available for
star-formation at lower redshifts.  How might this come about?  One
scenario involves the transition of gas from the warm to cold phase,
which occurs at a density $\sim
0.1$--$1\,\mathrm{cm}^{-3}$ \citep[e.g.,][]{Wolfire95,Wolfire03}.  As
the gas density decreases towards lower redshift, a
progressively smaller mass fraction of the neutral gas phase is cold.  
Additionally, the nature of gas collapse in the disk is critical, as the
galaxy-wide star formation rate may be particularly sensitive to
whether the central gaseous structure (typically a clumpy disk) is
violently self-gravitating \citep{dekel09,ceverino10}.  Star formation
is most efficient when large gas clumps become self-gravitating, and
large self-gravitating clumps of gas have been observed to form stars
with high efficiency at $z \sim$ 3 \citep[e.g.,][]{genzel11}.

Another possibility is that
we are witnessing the effect of the growth of the circumgalactic
medium (CGM) around galaxies.  Observations with the Cosmic Origins
Spectrograph on {\it HST} have recently revealed circumgalactic
ionized gas halos at low redshift containing significant baryon and
synthesized metal fractions \citep{werk14}.  The CGM likely grew over
time, as shock waves driven by supernova blast waves and superbubbles
raised an increasing cumulative gas fraction to high temperatures.
This reprocessed gas would be deposited in a warm ionized CGM and, at
least for a period of time, off limits to conversion into the cold
phase.  However, the physics of the CGM is poorly understood, and in
particular it is not clear at what rate the CGM gas recondenses
back into the warm neutral phase in the disk.

Finally, the feedback which builds the CGM may be less efficient
at high redshift, further increasing the amount of cold gas available
for star formation.  One may expect that the effects of feedback are less significant at
high redshift where halos are denser and at fixed mass have higher
circular velocities.  Both make it more difficult for SNe to power
galactic outflows that eject material into the CGM or outside the
halo.  Using the virial estimate from \citet{bryan98} we calculated
the circular velocities of the host halos
in our galaxy sample to be 242, 219, 219
and 197 km s$^{-1}$ at $z =$ 4, 5, 6 and 7, respectively.  Thus, the circular
velocities in fact \emph{decrease} towards higher redshift, which
would if anything make it easier to eject material at higher redshift.  
However, the properties of the feedback
mechanisms may evolve as well.
First, as galaxies evolve and the typical gas density
decreases, the supernova remnant thermalization efficiency, which is
an increasing function of the cooling time in \ion{H}{2} regions
shock-heated by supernova blastwaves, increases, i.e., a progressively
smaller fraction of the initial $\sim10^{51}\,\mathrm{erg}$ per
supernova is quickly irradiated \citep[see, e.g.,][]{creasey13}.  This
implies that at lower redshifts, a larger fraction of supernova energy
may be available to drive galaxy-wide outflows, thus making the
feedback from star formation more effective. It is also possible that
due to the very short growth times at high redshifts, so much gas is
piling on that the outflows are somehow bottled in. Furthermore, 
the higher dust content at lower redshift can lead to stronger momentum-driven radiative
stellar feedback, regulating further star formation \citep[see, e.g.,][]{murray10,andrews11}.

One additional potential physical mechanism which may evolve is the
ability of feedback from active galactic nuclei (AGNs) to suppress
star formation, which is commonly implemented in
theoretical models to avoid an overabundance of bright/massive
galaxies \citep[see discussion in][]{somerville14}.  This type of
feedback requires an accreting supermassive black hole, and although
there are some examples of bright AGNs at very high redshift, the
AGN/quasar luminosity function appears to decrease rapidly at $z >$ 3
\citep[e.g.,][]{richards06}.  \citet{bowler15} have recently observed that the bright end of the
galaxy UV luminosity function was steeper than the halo mass function
at $z =$ 6, but not at $z =$ 7, and hypothesized that such an observation could be
explained if feedback in bright/massive galaxies due to AGN first
``turned on'' at $z \lesssim$ 6.  However, the details of how AGN
couple with galaxies and their surroundings, particularly at these
epochs, are highly uncertain, so it remains unclear whether black hole
accretion has significantly affected the growth of the galaxies we
consider here.

While the scenarios we have discussed are clearly speculative, our observations imply that
the latter effects, primarily a reduced efficiency of feedback at
higher redshift due to
a variety of redshift-dependent effects, control the evolution of the
stellar baryon fraction.  We
conclude that the true cause of how a larger
fraction of the baryons turns into stars at higher redshifts is most
certainly determined by a delicate competition of factors.

Regardless of the underlying cause, the consequences of the increased
availability of cold gas are intriguing.
Our results at $z = 6$ and 7 show that $\sim$10\%--12\% of
the cosmic complement of baryons in these galaxies has been converted into stars.  
The remaining baryons must exist in the gas
phase.  If they are in the warm or cold neutral phase or the molecular phase, then the gas fraction in these phases is much
higher than at lower redshifts \citep{magdis12}.
A high gas mass fraction at very high redshifts is
not unexpected
and may soon be confirmed 
the Atacama Large Millimeter Array (ALMA; e.g., by measuring
the dynamical mass via spectrally resolved FIR emission lines).
There are indirect empirical hints that neutral and molecular gas fractions increase with redshift.
\citet{papovich11} studied the evolution of observed galaxy SFRs,
stellar masses, and sizes, and
concluded that the gas-to-stellar mass fraction must rise with
redshift, reaching $M_{\rm gas}/M_{\ast} = 3.9$ at $z =7$ (at a higher cumulative number density of 
$2\times10^{-4}$ Mpc$^{-3}$).  This could 
intriguingly play
some role in the decreasing visibility of Ly$\alpha$ at $z >6$  \citep[e.g.,][]{fontana10,pentericci11,tilvi14,pentericci14}.

Finally, we examine the likely descendants of these bright $z =$ 4--7 galaxies.
Specifically, being among the most rapidly star forming galaxies at their
redshifts, could some of them end up evolving into extreme systems such as
sub-millimeter galaxies (SMGs) by lower redshifts?  Using the halo mass
evolution tool from \citet{behroozi13c}, we calculated the 68\%
confidence range of the descendant halo masses of our sample of bright
high-redshift galaxies.  We find that by $z =$ 2, galaxies we observe
at $z =$ 4, 5, 6, and 7 will exist in halos with
$\log(M_\mathrm{h}/M_\odot) =$ 12.3--12.8, 12.3--12.8, 12.4--13, and
12.4--13.1, respectively.  SMGs are thought to be hosted by halos with
$\log(M_\mathrm{h}/M_\odot) \approx 13$ \citep[e.g.,][]{hickox12,casey14}, thus the majority of 
$M^{\ast}_\mathrm{UV}$ galaxies at $z =$ 4--5 will evolve into
galaxies at $z = 2$ with halos slightly less massive than those of the typical
SMGs.  However, the SMG host halo mass
begins to be consistent with the expected descendants of the $z =$ 6--7 galaxies, thus some
subset of very UV-bright galaxies at $z >6$ may, in principle, turn into
lower-redshift SMGs.  A basic test of this is to see whether, if we
assume these galaxies keep their current SFRs, they can grow large
enough to match the stellar mass of a typical SMG by $z =2$.  If we
assume a SFR of $50\,M_\odot$ yr$^{-1}$ (Table 1), we find that the $z =2$
descendants of $M^{\ast}_\mathrm{UV}$ galaxies at $z =6$ and 7 will
have $\log(M_\ast/M_\odot) = 11.1$. This is in the range of
SMG stellar masses found in the literature \citep[see review in
][]{casey14}.  Our assumed constant SFR of $\sim50\,
M$\sol\ yr$^{-1}$, which implies stellar masses consistent with those observed in SMGs,
is approximately the SFR disk galaxies should have prior to
coalescence to produce SMGs in the merger-driven scenario for SMG formation \citep{narayanan10}.

\section{Conclusions}

Recent observations have shown that the characteristic UV luminosity
$M^{\ast}_{\rm UV}$ does not significantly evolve from $z =$ 4 to 7,
which is unexpected given the general decline in the cosmic SFR
density towards higher redshift over that time.  To investigate the
physical effects behind this observed non-evolution in $M^{\ast}_{\rm
UV}$, we have inspected the stellar populations in $M^{\ast}_{\rm
UV}$ galaxies at $z =$ 4 to 7.  We have found the following results:

\begin{itemize}

\item Galaxies with $M_{\rm  UV} < -$ 21 appear to have very similar
physical properties at $z =$ 4, 5, 6, and 7, including stellar mass,
dust attenuation and SFR.

\item Using abundance matching to infer the likely hosting halo
masses, we found that $M_{\rm  UV} < -21$ galaxies, which we have
measured to have $\log(M_{\ast}/M_\odot)=$9.6--9.9, live in
progressively smaller halos towards higher redshift, with
$\log(M_\mathrm{h}/M_\odot) =$ 11.93 at $z =$ 4, and
$\log(M_\mathrm{h}/M_\odot) =$ 11.35 at $z =$ 7.

\item The stellar baryon fraction, defined as the fraction of baryons
in stars in units of the cosmic mean $\Omega_{\rm b}/\Omega_{\rm m}$
rises from 0.051 $\pm 0.006$ at $z =4$  to 0.117 $\pm 0.043$ at $z
= 7$.  This trend is significant at the 3.2$\sigma$ level.

\item The observed trend of an increasing SBF with increasing redshift
does not agree with simple expectations of how galaxies grow.  
Therefore, our observations imply a change in the
physical properties governing star-formation at $z >4$, such as, for example, a reduced efficiency of stellar and supernova feedback toward higher redshifts.

\end{itemize}

Future studies can improve upon our results by probing a larger
volume to increase the sample of bright galaxies, allowing us to
establish the evolution of the SBF at greater significance, as well as
extending this analysis to $z =$ 8 and 9.  Additionally, a more robust
determination of the ratio between stellar mass and halo mass, and
thus the SBF, can be done with accurate stellar mass functions,
which are only now being computed at $z \geq 6$.  Finally, through ALMA followup of
distant galaxies, we will begin to not only directly probe their dust
emission, removing some of the potential systematic biases inherent
when assuming a dust attenuation curve, but ALMA can also directly probe the evolution of galaxy gas
reservoirs with redshift. A direct observation of
increasing cold gas reservoirs with increasing redshift would provide
a complementary observation pointing to decreased feedback at high
redshift leading to an increased stellar baryon fraction.

\acknowledgements
We thank Caitlin Casey for useful discussions on the abundance of
dusty galaxies at high redshift.  SLF acknowledges support from the University of Texas
at Austin College of Natural Sciences.  SLF and MS were also supported by a NASA
Astrophysics and Data Analysis Program
award issued by JPL/Caltech.  MM acknowledges support from NSF grant AST-1413501.
Partial support for DN was provided by
NSF grants AST-1009452, AST-1442650, NASA grant HST-AR-13906.001.
This work is based on observations made with the NASA/ESA Hubble Space Telescope,
obtained at the Space Telescope Science
Institute, which is operated by the Association of Universities for
Research in Astronomy, Inc., under NASA contract NAS 5-26555. These
observations are associated with program \#12060.
This work is also based in part on
observations made with the Spitzer Space Telescope, which is operated
by the Jet Propulsion Laboratory, California Institute of Technology
under a contract with NASA. Support for this work was provided by NASA
through an award issued by JPL/Caltech.


\appendix
\setcounter{table}{0}
\renewcommand{\thetable}{A\arabic{table}}

In the above text, we discuss the median stellar population properties
of our sample of galaxies.  Galaxies of course are an inhomogeneous
population, thus in this appendix we provide tables of the individual
properties for each of the bright galaxies used in our analysis (those
with IRAC measurements, and a best-fitting SED model with $\chi^2 <$
20; see \S 2.3).
Each column in these tables lists the best-fitting value along with
the 68\% confidence range in parentheses.

\LongTables
\begin{deluxetable*}{cccccccc}
\tabletypesize{\small}
\tablecaption{Stellar Populations of Bright Galaxies at $z =$ 4}
\tablewidth{0pt}
\tablehead{
\colhead{ID} & \colhead{Right Ascension} & \colhead{Declination} & \colhead{Redshift} & \colhead{log M$_{\ast}$} & \colhead{Age} & \colhead{E(B-V)} & \colhead{SFR}\\
\colhead{$ $} & \colhead{(J2000)} & \colhead{(J2000)} & \colhead{$ $} & \colhead{(M\sol)} & \colhead{(Myr)} & \colhead{$ $} & \colhead{(M\sol\ yr$^{-1}$)}\\
}
\startdata
z4\_GSD\_34736&53.096840&-27.866074&3.51 (3.44--3.66)&9.97 (10.13--10.49)&10 (19--101)&0.34 (0.18--0.34)&451 (109--434)\\
z4\_GSD\_30292&53.086891&-27.844139&3.51 (3.46--3.75)&10.49 (10.31--10.51)&57 (49--90)&0.24 (0.16--0.32)&132 (64--322)\\
z4\_GNW\_9013&189.085114&62.160465&3.54 (3.49--3.60)&9.97 (9.93--10.15)&1015 (202--1015)&0.10 (0.02--0.10)&54 (26--53)\\
z4\_ERS\_4095&53.143330&-27.690090&3.55 (3.48--3.67)&10.27 (10.17--10.55)&49 (30--101)&0.16 (0.14--0.28)&88 (73--299)\\
z4\_ERS\_9969&53.120926&-27.709446&3.57 (3.52--3.64)&9.68 (9.43--9.68)&57 (30--286)&0.04 (0.00--0.10)&20 (14--38)\\
z4\_GNW\_26176&189.483200&62.284786&3.63 (3.57--3.71)&9.99 (9.96--10.18)&90 (40--202)&0.12 (0.02--0.14)&65 (27--85)\\
z4\_GNW\_1986&189.159058&62.115471&3.64 (3.58--3.70)&9.98 (9.96--10.19)&101 (49--1015)&0.08 (0.06--0.10)&55 (44--77)\\
z4\_ERS\_16929&53.087231&-27.729538&3.64 (3.58--3.72)&9.42 (9.21--9.85)&19 (10--40)&0.14 (0.14--0.20)&59 (55--109)\\
z4\_GSD\_29028&53.087368&-27.839535&3.64 (3.52--3.75)&10.17 (10.12--10.22)&80 (80--101)&0.10 (0.06--0.10)&38 (25--40)\\
z4\_ERS\_20075&53.020580&-27.742151&3.65 (3.59--3.71)&10.55 (10.54--10.66)&202 (202--570)&0.06 (0.02--0.08)&75 (53--104)\\
z4\_GND\_30689&189.339355&62.216450&3.66 (3.59--3.75)&9.45 (9.40--9.70)&39 (30--71)&0.04 (0.00--0.10)&20 (14--36)\\
z4\_GSW\_4356&53.109478&-27.879360&3.67 (3.61--3.77)&9.72 (9.66--9.90)&57 (49--90)&0.00 (0.00--0.08)&19 (19--43)\\
z4\_GSD\_15786&53.071735&-27.798437&3.67 (3.60--3.75)&9.61 (9.67--9.88)&49 (57--1015)&0.04 (0.00--0.10)&19 (14--33)\\
z4\_GSD\_535&53.198959&-27.737940&3.69 (3.57--3.81)&9.59 (9.53--9.79)&49 (40--57)&0.00 (0.00--0.08)&18 (17--39)\\
z4\_ERS\_3396&53.117710&-27.686771&3.69 (3.64--3.75)&9.90 (9.53--9.90)&49 (30--49)&0.10 (0.08--0.16)&48 (40--74)\\
z4\_GSD\_21002&53.121414&-27.814621&3.69 (3.63--3.76)&9.80 (9.57--9.85)&30 (10--30)&0.18 (0.16--0.24)&93 (70--152)\\
z4\_GSD\_35257&53.107422&-27.869299&3.70 (3.58--3.82)&10.46 (10.24--10.47)&101 (71--101)&0.28 (0.10--0.28)&222 (45--240)\\
z4\_ERS\_11888&53.069328&-27.714815&3.70 (3.62--3.78)&9.55 (9.29--9.76)&49 (19--49)&0.00 (0.00--0.18)&14 (14--77)\\
z4\_GSD\_11269&53.031239&-27.785215&3.71 (3.61--3.79)&10.06 (10.02--10.16)&49 (40--101)&0.10 (0.10--0.24)&45 (43--176)\\
z4\_GSD\_27735&53.138859&-27.835371&3.72 (3.65--3.79)&9.56 (9.36--9.63)&30 (19--57)&0.14 (0.02--0.18)&54 (17--78)\\
z4\_GSW\_2898&53.144775&-27.871527&3.78 (3.68--3.87)&10.18 (9.88--10.20)&39 (30--49)&0.20 (0.06--0.26)&131 (35--200)\\
z4\_GSD\_31543&53.066261&-27.849056&3.81 (3.75--3.98)&10.12 (10.00--10.20)&49 (30--57)&0.12 (0.10--0.20)&52 (41--117)\\
z4\_GND\_38889&189.181412&62.189281&3.81 (3.75--4.05)&10.40 (10.38--10.48)&57 (49--71)&0.14 (0.14--0.22)&68 (65--163)\\
z4\_ERS\_4079&53.110340&-27.689985&3.81 (3.74--3.99)&10.09 (9.87--10.17)&19 (10--57)&0.32 (0.06--0.34)&325 (31--373)\\
z4\_GSW\_7015&53.189873&-27.892590&3.83 (3.75--3.93)&10.15 (9.77--10.15)&49 (30--80)&0.16 (0.06--0.22)&84 (32--152)\\
z4\_GSW\_6936&53.073589&-27.892235&3.83 (3.76--3.92)&10.47 (10.40--10.49)&286 (202--718)&0.00 (0.00--0.12)&22 (22--81)\\
z4\_GSW\_9851&53.181850&-27.906641&3.85 (3.79--3.95)&9.97 (9.64--9.94)&49 (30--49)&0.12 (0.00--0.20)&56 (18--114)\\
z4\_GNW\_12987&189.040390&62.186352&3.92 (3.84--4.00)&9.66 (9.49--9.69)&57 (40--71)&0.02 (0.00--0.02)&19 (15--23)\\
z4\_ERS\_5026&53.133690&-27.693453&3.93 (3.82--4.05)&9.67 (9.63--9.90)&39 (30--101)&0.06 (0.00--0.12)&27 (16--49)\\
z4\_GSD\_34857&53.076183&-27.866360&3.98 (3.89--4.11)&10.32 (10.23--10.37)&286 (40--286)&0.16 (0.14--0.16)&123 (96--128)\\
z4\_GSD\_23593&53.232452&-27.822868&3.99 (3.89--4.14)&9.48 (9.33--9.92)&19 (19--80)&0.18 (0.14--0.22)&72 (52--101)\\
z4\_GSD\_905&53.168266&-27.741940&4.01 (3.92--4.11)&9.78 (9.44--9.76)&90 (19--90)&0.00 (0.00--0.20)&13 (13--88)\\
z4\_GNW\_25081&189.331039&62.290836&4.01 (3.90--4.10)&9.39 (9.34--9.43)&10 (10--10)&0.26 (0.24--0.26)&140 (122--144)\\
z4\_ERS\_3543&53.144386&-27.687588&4.02 (3.87--4.16)&10.01 (9.96--10.26)&39 (40--404)&0.16 (0.06--0.16)&78 (30--83)\\
z4\_GND\_40720&189.179291&62.182003&4.02 (3.84--4.20)&9.70 (9.26--9.74)&30 (19--80)&0.20 (0.02--0.20)&75 (12--83)\\
z4\_GNW\_18460&189.286835&62.367325&4.05 (3.92--4.16)&10.28 (10.22--10.38)&286 (202--806)&0.00 (0.00--0.10)&20 (19--53)\\
z4\_ERS\_22264&53.075272&-27.755194&4.06 (3.96--4.17)&10.12 (10.01--10.24)&39 (30--508)&0.16 (0.10--0.16)&93 (62--99)\\
z4\_GND\_27047&189.356354&62.227554&4.10 (3.92--4.21)&10.47 (10.26--10.54)&202 (49--904)&0.12 (0.12--0.20)&52 (50--98)\\
z4\_GNW\_21799&189.329178&62.331532&4.10 (4.03--4.17)&9.82 (9.69--9.96)&57 (40--80)&0.00 (0.00--0.04)&24 (23--35)\\
z4\_GNW\_23907&189.467804&62.297764&4.10 (3.98--4.21)&9.94 (9.82--10.31)&80 (19--101)&0.20 (0.14--0.24)&122 (79--164)\\
z4\_GND\_7728&189.270905&62.291943&4.10 (4.01--4.20)&10.23 (9.78--10.24)&101 (19--101)&0.20 (0.16--0.22)&133 (91--163)\\
z4\_GSD\_15152&53.027557&-27.796583&4.12 (4.05--4.22)&10.41 (10.23--10.44)&101 (49--101)&0.20 (0.20--0.24)&171 (178--266)\\
z4\_GNW\_17778&189.311584&62.382095&4.14 (4.00--4.29)&9.76 (9.29--9.82)&30 (19--49)&0.18 (0.10--0.20)&85 (41--99)\\
z4\_ERS\_11615&53.107601&-27.713976&4.14 (4.01--4.26)&9.52 (9.61--9.96)&19 (30--101)&0.18 (0.02--0.18)&79 (18--83)\\
z4\_GND\_23790&189.233093&62.236786&4.15 (4.06--4.27)&10.50 (10.49--10.60)&101 (57--404)&0.18 (0.08--0.18)&211 (83--221)\\
z4\_GSD\_20508&53.192692&-27.813051&4.17 (4.06--4.26)&9.65 (9.61--9.92)&30 (30--202)&0.14 (0.04--0.16)&59 (25--76)\\
z4\_GNW\_18340&189.299255&62.370079&4.19 (4.03--4.30)&11.29 (11.17--11.33)&1015 (904--1015)&0.18 (0.18--0.24)&146 (142--226)\\
z4\_GNW\_15232&189.014038&62.200378&4.22 (4.11--4.31)&10.18 (9.80--10.18)&71 (30--80)&0.22 (0.20--0.26)&178 (139--242)\\
z4\_ERS\_14762&53.016903&-27.723013&4.22 (4.11--4.32)&9.40 (9.32--9.84)&19 (19--80)&0.16 (0.08--0.18)&57 (28--78)\\
z4\_GNW\_11056&189.013092&62.173153&4.23 (4.08--4.31)&9.64 (9.61--9.78)&39 (30--57)&0.04 (0.00--0.08)&26 (16--38)\\
z4\_GSW\_66&53.230198&-27.839573&4.24 (4.17--4.33)&9.76 (9.76--10.22)&19 (19--90)&0.18 (0.12--0.18)&138 (80--143)\\
z4\_GSD\_9138&53.215435&-27.778782&4.24 (4.10--4.37)&10.07 (9.82--10.08)&101 (80--101)&0.18 (0.00--0.20)&92 (18--102)\\
z4\_GNW\_3896&189.145218&62.129818&4.24 (4.12--4.39)&9.80 (9.77--10.12)&10 (10--40)&0.34 (0.32--0.34)&346 (297--359)\\
z4\_GNW\_8115&189.145187&62.154770&4.25 (4.09--4.34)&9.62 (9.54--9.79)&49 (40--80)&0.02 (0.00--0.06)&19 (15--31)\\
z4\_GNW\_26546&189.445068&62.281898&4.26 (4.02--4.38)&10.36 (10.35--10.59)&101 (57--286)&0.28 (0.00--0.30)&151 (12--228)\\
z4\_GND\_9704&189.201706&62.278599&4.27 (4.19--4.36)&9.85 (9.85--10.23)&30 (30--90)&0.14 (0.04--0.16)&95 (38--117)\\
z4\_GND\_21065&189.290955&62.244240&4.29 (4.13--4.41)&10.47 (10.17--10.53)&202 (80--286)&0.30 (0.18--0.30)&264 (81--240)\\
z4\_GND\_31952&189.119812&62.212612&4.29 (4.19--4.37)&9.56 (9.50--9.90)&30 (30--57)&0.12 (0.06--0.14)&45 (27--58)\\
z4\_GNW\_24624&189.388474&62.294460&4.30 (4.17--4.43)&9.34 (9.35--9.62)&39 (30--57)&0.00 (0.00--0.10)&12 (11--41)\\
z4\_GNW\_11698&188.988419&62.177494&4.30 (4.19--4.39)&9.62 (9.42--9.66)&30 (19--49)&0.16 (0.02--0.20)&61 (14--82)\\
z4\_GNW\_460&189.120911&62.101513&4.30 (4.18--4.40)&9.41 (9.36--9.80)&30 (19--71)&0.08 (0.00--0.16)&27 (14--64)\\
z4\_GNW\_2957&189.185181&62.122967&4.30 (4.17--4.42)&9.90 (9.89--10.12)&30 (30--49)&0.16 (0.12--0.18)&84 (59--95)\\
z4\_GNW\_31531&189.295654&62.349960&4.31 (4.15--4.43)&10.32 (10.10--10.46)&404 (49--286)&0.24 (0.16--0.30)&122 (54--208)\\
z4\_GNW\_7377&189.108002&62.151005&4.31 (4.25--4.37)&10.35 (10.24--10.35)&80 (30--80)&0.00 (0.00--0.12)&57 (56--179)\\
z4\_GSD\_21252&53.143112&-27.815502&4.33 (4.24--4.39)&10.05 (10.03--10.47)&57 (30--101)&0.18 (0.16--0.22)&181 (149--266)\\
z4\_GND\_30347&189.324326&62.217796&4.34 (4.24--4.41)&9.71 (9.54--9.82)&1015 (40--404)&0.08 (0.06--0.12)&26 (21--39)\\
z4\_GNW\_24183&189.466507&62.296875&4.35 (4.21--4.43)&10.09 (9.95--10.22)&570 (71--570)&0.14 (0.10--0.18)&68 (42--107)\\
z4\_GNW\_21312&189.366592&62.330898&4.38 (4.25--4.46)&9.66 (9.72--10.34)&10 (10--30)&0.26 (0.22--0.32)&210 (154--394)\\
z4\_GND\_27301&189.093216&62.226814&4.39 (4.29--4.47)&10.00 (9.90--10.14)&904 (40--718)&0.14 (0.10--0.18)&50 (37--74)\\
z4\_GSD\_16522&53.076065&-27.800694&4.39 (4.32--4.47)&9.67 (9.52--10.14)&19 (10--90)&0.22 (0.22--0.26)&160 (154--224)\\
z4\_GNW\_21830&189.366791&62.331589&4.40 (4.30--4.48)&9.95 (9.95--10.26)&30 (30--286)&0.20 (0.14--0.20)&118 (65--120)\\
z4\_GND\_7158&189.252747&62.298897&4.41 (4.31--4.50)&10.27 (9.94--10.30)&101 (40--101)&0.26 (0.22--0.28)&146 (90--167)\\
z4\_GNW\_32075&189.304321&62.353088&4.41 (4.31--4.48)&10.46 (10.45--10.77)&30 (30--202)&0.26 (0.20--0.26)&384 (229--397)\\
z4\_GNW\_7816&189.049530&62.152920&4.42 (4.33--4.49)&9.70 (9.71--10.15)&19 (19--71)&0.18 (0.12--0.20)&100 (64--121)\\
z4\_GSW\_5453&53.199787&-27.884937&4.42 (4.32--4.49)&9.90 (9.77--10.03)&286 (80--570)&0.16 (0.14--0.20)&52 (42--75)\\
z4\_GND\_39360&189.242157&62.187439&4.44 (4.32--4.54)&9.86 (9.79--10.15)&71 (30--90)&0.18 (0.16--0.24)&65 (54--118)\\
z4\_GSW\_1319&53.201015&-27.860250&4.44 (4.37--4.51)&9.84 (9.60--9.87)&39 (30--90)&0.14 (0.00--0.18)&60 (16--89)\\
z4\_GNW\_8416&189.195557&62.156864&4.45 (4.31--4.54)&10.51 (10.39--10.58)&286 (49--286)&0.24 (0.20--0.30)&136 (91--259)\\
z4\_GND\_32575&189.173126&62.210808&4.45 (4.36--4.53)&9.75 (9.37--9.72)&39 (30--49)&0.14 (0.06--0.18)&49 (21--72)\\
z4\_GNW\_13552&189.061249&62.189472&4.45 (4.34--4.54)&9.92 (9.94--10.38)&19 (19--71)&0.24 (0.20--0.26)&164 (109--184)\\
z4\_GNW\_8070&189.107422&62.154560&4.45 (4.37--4.53)&9.81 (9.71--9.88)&30 (30--49)&0.16 (0.04--0.18)&96 (29--105)\\
z4\_ERS\_5818&53.069077&-27.696518&4.47 (4.38--4.56)&9.43 (9.37--9.64)&39 (30--49)&0.02 (0.00--0.16)&16 (13--65)\\
z4\_GNW\_20572&189.313507&62.320702&4.47 (4.37--4.56)&9.83 (9.52--9.86)&39 (30--90)&0.16 (0.04--0.16)&58 (17--56)\\
z4\_GNW\_18301&189.335205&62.370796&4.47 (4.39--4.54)&9.97 (9.88--10.31)&30 (30--90)&0.18 (0.16--0.22)&116 (98--194)\\
z4\_GNW\_18575&189.297821&62.365429&4.49 (4.42--4.55)&10.70 (10.37--10.74)&30 (19--57)&0.28 (0.24--0.30)&739 (458--894)\\
z4\_GNW\_2261&189.192337&62.117081&4.50 (4.42--4.57)&9.96 (9.62--9.97)&39 (30--80)&0.14 (0.02--0.18)&79 (23--115)\\
z4\_GND\_40010&189.335739&62.184937&4.51 (4.44--4.56)&10.16 (9.98--10.20)&57 (30--90)&0.06 (0.00--0.16)&62 (35--158)\\
z4\_GNW\_7213&189.108322&62.150093&4.51 (0.69--4.57)&9.89 (6.51--9.87)&30 (19--90)&0.20 (0.10--0.66)&116 (0--142)\\
z4\_GNW\_7206&189.108185&62.149807&4.54 (4.46--4.61)&9.97 (9.85--10.12)&19 (19--30)&0.26 (0.22--0.28)&245 (149--279)\\
z4\_GSW\_8512&53.177471&-27.900093&4.54 (4.48--4.60)&9.53 (9.54--10.02)&19 (19--40)&0.18 (0.18--0.26)&67 (67--152)\\
z4\_GND\_25942&189.147903&62.230583&4.54 (4.44--4.63)&10.09 (9.89--10.23)&19 (19--57)&0.30 (0.24--0.32)&325 (180--360)\\
z4\_GSD\_36028&53.079254&-27.877260&4.54 (4.46--4.59)&10.21 (9.96--10.22)&1015 (30--202)&0.10 (0.06--0.12)&83 (56--104)\\
z4\_GND\_14271&189.305084&62.263287&4.56 (4.44--4.65)&10.79 (10.45--10.79)&39 (40--80)&0.36 (0.30--0.36)&537 (295--541)\\
z4\_GNW\_18613&189.276077&62.364826&4.57 (4.47--4.64)&10.04 (10.06--10.58)&19 (19--57)&0.22 (0.20--0.26)&219 (184--335)\\
\enddata
\end{deluxetable*}

\clearpage

\LongTables
\begin{deluxetable*}{cccccccc}
\tabletypesize{\small}
\tablecaption{Stellar Populations of Bright Galaxies at $z =$ 5}
\tablewidth{0pt}
\tablehead{
\colhead{ID} & \colhead{Right Ascension} & \colhead{Declination} & \colhead{Redshift} & \colhead{log M$_{\ast}$} & \colhead{Age} & \colhead{E(B-V)} & \colhead{SFR}\\
\colhead{$ $} & \colhead{(J2000)} & \colhead{(J2000)} & \colhead{$ $} & \colhead{(M\sol)} & \colhead{(Myr)} & \colhead{$ $} & \colhead{(M\sol\ yr$^{-1}$)}\\
}
\startdata
z5\_GSW\_8762&53.208008&-27.901289&4.51 (4.44--4.57)&9.72 (9.50--9.78)&49 (30--71)&0.06 (0.00--0.08)&31 (16--38)\\
z5\_GND\_36639&189.186142&62.197327&4.56 (4.45--4.68)&10.25 (10.11--10.52)&80 (19--101)&0.28 (0.26--0.32)&223 (203--330)\\
z5\_GNW\_17976&189.300125&62.377483&4.57 (4.42--4.74)&10.19 (10.12--10.37)&286 (80--286)&0.22 (0.10--0.26)&102 (31--164)\\
z5\_GSD\_9044&53.091724&-27.778580&4.58 (4.48--4.67)&10.26 (10.02--10.25)&202 (40--508)&0.06 (0.06--0.18)&32 (33--110)\\
z5\_GSW\_6918&53.234589&-27.892109&4.66 (4.58--4.75)&9.52 (9.38--9.59)&57 (49--57)&0.00 (0.00--0.02)&14 (13--18)\\
z5\_GND\_38041&189.299210&62.192570&4.67 (4.60--4.74)&9.36 (9.31--9.89)&10 (10--19)&0.22 (0.20--0.28)&106 (94--198)\\
z5\_GND\_16758&189.176086&62.256329&4.71 (4.62--4.84)&10.03 (9.96--10.11)&19 (19--30)&0.26 (0.26--0.28)&277 (272--336)\\
z5\_GSD\_33149&53.070778&-27.856453&4.71 (4.62--4.81)&10.02 (9.63--10.30)&101 (10--508)&0.20 (0.10--0.24)&116 (49--172)\\
z5\_GSD\_13326&53.095345&-27.790989&4.71 (4.64--4.79)&10.19 (9.98--10.26)&30 (19--80)&0.26 (0.16--0.26)&228 (86--238)\\
z5\_ERS\_3475&53.070839&-27.687143&4.72 (4.66--4.81)&9.95 (9.84--10.39)&19 (19--40)&0.26 (0.26--0.32)&205 (211--369)\\
z5\_GND\_38212&189.273590&62.192028&4.72 (4.65--5.06)&9.84 (9.82--10.19)&30 (30--71)&0.16 (0.08--0.20)&72 (43--136)\\
z5\_GSW\_1565&53.238213&-27.862486&4.75 (4.70--4.79)&9.96 (9.51--9.96)&30 (19--30)&0.18 (0.12--0.20)&135 (73--158)\\
z5\_GND\_12253&189.294418&62.269447&4.77 (4.70--4.90)&10.06 (9.87--10.19)&570 (40--718)&0.14 (0.10--0.20)&55 (39--104)\\
z5\_ERS\_12604&53.021912&-27.716784&4.81 (4.76--4.89)&9.60 (9.60--9.84)&39 (40--1015)&0.08 (0.02--0.10)&30 (17--34)\\
z5\_GNW\_25539&189.489624&62.288536&4.81 (4.76--4.87)&10.34 (10.03--10.43)&19 (19--40)&0.26 (0.18--0.26)&570 (266--567)\\
z5\_ERS\_2517&53.119019&-27.682158&4.82 (4.76--4.93)&9.70 (9.41--9.77)&30 (30--80)&0.18 (0.00--0.16)&73 (14--68)\\
z5\_GND\_17343&189.091400&62.254662&4.83 (4.75--4.95)&10.16 (10.05--10.19)&30 (19--71)&0.24 (0.06--0.24)&214 (45--230)\\
z5\_ERS\_6044&53.048820&-27.697111&4.85 (4.78--4.94)&9.85 (9.78--9.87)&90 (30--90)&0.00 (0.00--0.20)&15 (15--94)\\
z5\_GSW\_6966&53.245884&-27.892273&4.94 (0.83--4.98)&10.29 (10.20--10.37)&19 (10--30)&0.34 (0.28--0.36)&506 (265--566)\\
z5\_GNW\_6112&189.064835&62.143963&4.95 (4.90--5.06)&9.85 (9.47--10.02)&19 (10--71)&0.26 (0.06--0.28)&183 (31--221)\\
z5\_GSD\_10352&53.021172&-27.782366&4.95 (4.91--4.99)&10.43 (10.36--10.80)&57 (19--90)&0.22 (0.08--0.24)&474 (137--571)\\
z5\_GNW\_4779&189.203064&62.136204&4.96 (4.88--5.15)&9.86 (10.07--10.31)&19 (30--286)&0.28 (0.04--0.32)&174 (20--244)\\
z5\_GND\_15230&189.205124&62.260712&5.02 (4.96--5.08)&9.41 (9.41--9.69)&39 (30--49)&0.00 (0.00--0.12)&15 (15--59)\\
z5\_GNW\_16101&189.000168&62.207241&5.02 (4.78--5.06)&9.50 (9.20--9.52)&49 (30--49)&0.02 (0.00--0.04)&18 (12--20)\\
z5\_GNW\_10657&189.130219&62.170780&5.03 (4.80--5.15)&9.35 (9.19--9.44)&49 (30--57)&0.00 (0.00--0.06)&13 (11--27)\\
z5\_GNW\_19973&189.371780&62.324554&5.03 (4.80--5.12)&10.10 (10.15--10.59)&10 (10--202)&0.38 (0.22--0.34)&681 (140--473)\\
z5\_ERS\_2314&53.088764&-27.680889&5.05 (5.01--5.20)&9.55 (9.43--9.60)&30 (19--49)&0.14 (0.00--0.20)&52 (16--86)\\
z5\_PAR1\_1735&53.246998&-27.686445&5.07 (4.99--5.18)&10.06 (9.74--10.09)&49 (30--71)&0.16 (0.08--0.18)&69 (32--79)\\
z5\_GNW\_13254&189.039703&62.187725&5.24 (5.15--5.32)&9.88 (9.74--9.93)&71 (49--71)&0.00 (0.00--0.04)&22 (21--33)\\
z5\_ERS\_9511&53.155231&-27.700718&5.26 (5.05--5.40)&9.90 (9.49--9.99)&49 (19--90)&0.12 (0.00--0.20)&47 (14--98)\\
z5\_GNW\_25408&189.457428&62.289562&5.27 (5.11--5.40)&9.62 (9.35--9.66)&71 (40--80)&0.00 (0.00--0.02)&12 (11--15)\\
z5\_PAR1\_1385&53.273891&-27.683685&5.29 (5.20--5.36)&9.66 (9.67--9.99)&30 (30--71)&0.16 (0.12--0.22)&60 (42--113)\\
z5\_GSD\_17901&53.200577&-27.804909&5.33 (5.23--5.43)&9.91 (9.57--9.97)&57 (30--90)&0.08 (0.00--0.12)&34 (17--48)\\
z5\_GND\_34380&189.249146&62.205204&5.35 (5.24--5.45)&10.42 (10.35--10.61)&49 (40--508)&0.18 (0.14--0.24)&136 (85--234)\\
z5\_GND\_33094&189.266449&62.209164&5.36 (5.26--5.43)&9.82 (9.54--9.79)&49 (40--80)&0.10 (0.02--0.10)&39 (16--36)\\
z5\_GNW\_27194&189.523804&62.278839&5.40 (5.30--5.50)&10.14 (10.01--10.30)&57 (40--202)&0.08 (0.04--0.16)&50 (33--105)\\
z5\_GND\_39570&189.175720&62.186714&5.45 (5.33--5.55)&9.74 (9.80--10.22)&19 (30--49)&0.22 (0.12--0.22)&126 (49--140)\\
z5\_GND\_10054&189.169846&62.277298&5.46 (5.37--5.53)&9.92 (9.62--9.91)&57 (40--90)&0.08 (0.00--0.14)&35 (16--58)\\
z5\_GND\_35096&189.283508&62.203049&5.48 (5.35--5.55)&10.32 (9.81--10.38)&202 (19--101)&0.08 (0.04--0.22)&43 (31--167)\\
z5\_GSD\_4579&53.170231&-27.762848&5.48 (5.42--5.54)&9.67 (9.49--9.76)&30 (19--49)&0.12 (0.08--0.14)&49 (33--60)\\
z5\_GNW\_3960&189.055939&62.129990&5.52 (5.38--5.67)&9.88 (9.59--9.88)&71 (30--90)&0.02 (0.00--0.08)&23 (16--42)\\
z5\_GND\_18617&189.353439&62.250774&5.53 (5.46--5.63)&9.68 (9.56--9.91)&19 (19--30)&0.16 (0.12--0.18)&95 (67--122)\\
z5\_GND\_24948&189.320312&62.233444&5.54 (5.47--5.63)&9.53 (9.54--10.10)&10 (10--30)&0.26 (0.26--0.32)&179 (178--331)\\
z5\_GNW\_28218&189.503250&62.273884&5.55 (5.48--5.62)&10.47 (10.16--10.47)&30 (19--49)&0.26 (0.14--0.26)&435 (125--431)\\
z5\_GSD\_4436&53.170811&-27.762228&5.58 (5.49--5.65)&9.53 (9.53--9.78)&30 (19--49)&0.10 (0.02--0.18)&35 (17--84)\\
z5\_GNW\_29490&189.312943&62.344555&5.69 (5.55--5.84)&9.58 (9.58--9.89)&10 (10--19)&0.28 (0.28--0.30)&200 (182--254)\\
\enddata
\end{deluxetable*}

\clearpage

\LongTables
\begin{deluxetable*}{cccccccc}
\tabletypesize{\small}
\tablecaption{Stellar Populations of Bright Galaxies at $z =$ 6}
\tablewidth{0pt}
\tablehead{
\colhead{ID} & \colhead{Right Ascension} & \colhead{Declination} & \colhead{Redshift} & \colhead{log M$_{\ast}$} & \colhead{Age} & \colhead{E(B-V)} & \colhead{SFR}\\
\colhead{$ $} & \colhead{(J2000)} & \colhead{(J2000)} & \colhead{$ $} & \colhead{(M\sol)} & \colhead{(Myr)} & \colhead{$ $} & \colhead{(M\sol\ yr$^{-1}$)}\\
}
\startdata
z6\_GSD\_29074&53.156788&-27.839560&5.56 (5.46--5.66)&9.71 (9.54--10.02)&30 (19--90)&0.16 (0.00--0.18)&67 (16--85)\\
z6\_GSD\_17919&53.074165&-27.804928&5.65 (5.53--5.75)&9.72 (9.55--9.85)&80 (57--202)&0.00 (0.00--0.04)&13 (12--20)\\
z6\_GNW\_10970&189.075775&62.172729&5.67 (5.57--5.75)&10.52 (10.37--10.56)&404 (40--570)&0.20 (0.18--0.24)&170 (146--243)\\
z6\_GND\_16399&189.234833&62.257507&5.67 (5.58--5.76)&9.66 (9.44--9.75)&39 (19--57)&0.08 (0.02--0.18)&27 (16--87)\\
z6\_GND\_16819&189.328232&62.256317&5.69 (5.61--5.76)&10.09 (9.82--10.20)&19 (19--30)&0.28 (0.20--0.26)&319 (130--302)\\
z6\_ERS\_4104&53.066242&-27.689983&5.72 (5.61--5.84)&9.57 (9.48--9.67)&39 (30--49)&0.02 (0.00--0.04)&21 (17--27)\\
z6\_GSW\_12831&53.106689&-27.930193&5.79 (5.72--5.84)&10.05 (9.96--10.24)&49 (40--101)&0.06 (0.00--0.10)&57 (34--83)\\
z6\_GSW\_6659&53.151745&-27.890762&5.80 (5.68--5.91)&9.65 (9.39--9.82)&39 (19--71)&0.08 (0.00--0.14)&31 (14--52)\\
z6\_GSD\_27934&53.101673&-27.836084&5.80 (5.70--5.90)&10.61 (10.59--10.67)&19 (19--30)&0.38 (0.36--0.40)&1066 (863--1186)\\
z6\_MAIN\_5871&53.166721&-27.804167&5.80 (5.76--5.83)&9.90 (9.64--9.91)&57 (40--57)&0.04 (0.00--0.04)&34 (21--34)\\
z6\_GND\_28043&189.418427&62.224796&5.85 (5.75--5.94)&10.00 (9.69--10.14)&49 (19--80)&0.10 (0.02--0.18)&46 (25--104)\\
z6\_GNW\_22717&189.416199&62.333141&5.88 (5.66--6.04)&10.17 (9.74--10.27)&806 (30--806)&0.20 (0.02--0.22)&94 (17--117)\\
z6\_GSD\_23051&53.225368&-27.821125&5.96 (5.86--6.05)&10.07 (9.91--10.26)&39 (30--202)&0.14 (0.02--0.18)&89 (30--122)\\
z6\_GNW\_23437&189.388000&62.301167&6.02 (5.85--6.18)&9.84 (9.60--9.87)&101 (57--101)&0.00 (0.00--0.08)&12 (11--27)\\
z6\_PAR1\_1068&53.234592&-27.680861&6.13 (6.05--6.23)&10.10 (9.90--10.16)&39 (40--80)&0.18 (0.06--0.18)&109 (34--103)\\
z6\_GNW\_22555&189.301788&62.307808&6.15 (6.06--6.25)&9.63 (9.41--9.65)&57 (40--57)&0.02 (0.00--0.02)&18 (12--19)\\
z6\_ERS\_7413&53.158138&-27.702112&6.25 (6.12--6.39)&9.91 (9.66--10.07)&30 (19--49)&0.10 (0.00--0.14)&85 (38--127)\\
z6\_GSD\_21289&53.076241&-27.815453&6.30 (6.20--6.39)&9.63 (9.54--9.88)&39 (30--202)&0.10 (0.00--0.16)&32 (13--63)\\
z6\_GSD\_233&53.188583&-27.733210&6.39 (6.29--6.48)&9.88 (9.42--9.93)&30 (19--49)&0.20 (0.04--0.20)&112 (23--115)\\
\enddata
\end{deluxetable*}

\LongTables
\begin{deluxetable*}{cccccccc}
\tabletypesize{\small}
\tablecaption{Stellar Populations of Bright Galaxies at $z =$ 7}
\tablewidth{0pt}
\tablehead{
\colhead{ID} & \colhead{Right Ascension} & \colhead{Declination} & \colhead{Redshift} & \colhead{log M$_{\ast}$} & \colhead{Age} & \colhead{E(B-V)} & \colhead{SFR}\\
\colhead{$ $} & \colhead{(J2000)} & \colhead{(J2000)} & \colhead{$ $} & \colhead{(M\sol)} & \colhead{(Myr)} & \colhead{$ $} & \colhead{(M\sol\ yr$^{-1}$)}\\
}
\startdata
z7\_GSD\_25074&53.233047&-27.827383&6.66 (6.47--6.79)&10.09 (9.87--10.15)&404 (49--508)&0.18 (0.00--0.14)&82 (15--55)\\
z7\_ERS\_12574&53.094410&-27.716846&6.71 (6.58--6.85)&9.83 (9.63--10.06)&49 (30--202)&0.08 (0.00--0.18)&35 (16--90)\\
z7\_GNW\_24443&189.356888&62.295319&6.72 (6.65--6.81)&10.02 (9.80--10.05)&30 (30--71)&0.18 (0.00--0.18)&155 (29--160)\\
z7\_GNW\_24671&189.361710&62.294373&6.72 (6.27--7.07)&9.15 (8.53--9.25)&19 (10--40)&0.10 (0.00--0.10)&37 (11--33)\\
z7\_GSD\_21368&53.154922&-27.815744&6.75 (6.62--6.97)&10.14 (9.89--10.34)&404 (30--286)&0.20 (0.00--0.26)&82 (14--142)\\
z7\_GNW\_17001&189.032486&62.216415&6.76 (6.55--6.92)&10.74 (10.70--10.82)&404 (286--570)&0.06 (0.02--0.12)&22 (14--37)\\
z7\_GSD\_21172&53.155342&-27.815178&6.84 (6.76--6.96)&10.19 (9.95--10.20)&30 (19--80)&0.24 (0.00--0.30)&231 (24--396)\\
z7\_GSD\_10175&53.210335&-27.782211&7.05 (6.78--7.32)&9.70 (9.23--9.81)&19 (10--71)&0.22 (0.00--0.24)&129 (15--141)\\
z7\_GND\_18181&189.082687&62.252476&7.09 (6.94--7.28)&9.36 (9.28--9.96)&10 (10--49)&0.22 (0.12--0.24)&123 (47--142)\\
z7\_GND\_11402&189.186172&62.270863&7.16 (7.00--7.34)&9.69 (9.22--9.76)&30 (10--49)&0.10 (0.02--0.18)&51 (21--114)\\
z7\_GNW\_4703&189.094528&62.135540&7.17 (7.04--7.37)&9.55 (9.06--9.68)&39 (10--49)&0.02 (0.00--0.10)&27 (22--61)\\
z7\_GNW\_19939&189.273392&62.324783&7.25 (7.03--7.75)&8.79 (8.64--9.29)&10 (10--30)&0.10 (0.00--0.12)&34 (14--42)\\
z7\_GND\_42912&189.157883&62.302372&7.49 (7.33--7.70)&9.51 (9.41--9.65)&10 (10--10)&0.24 (0.20--0.28)&178 (133--224)\\
z7\_PAR2\_3098&53.281712&-27.867699&7.66 (7.44--7.82)&9.82 (9.56--9.90)&30 (19--49)&0.14 (0.02--0.18)&69 (23--111)\\
\enddata
\end{deluxetable*}

\end{document}